\documentclass[fleqn,usenatbib]{mnras}

\usepackage{newtxtext,newtxmath}
\usepackage[T1]{fontenc}

\DeclareRobustCommand{\VAN}[3]{#2}
\let\VANthebibliography\thebibliography
\def\thebibliography{\DeclareRobustCommand{\VAN}[3]{##3}\VANthebibliography}

\usepackage{graphicx}	% Including figure files
\usepackage{amsmath}	% Advanced maths commands
\usepackage{lscape}
\usepackage{orcidlink}
\usepackage{hyperref}

\newcommand{\rateunits}{\times 10^{4}\, \mathrm{yr}^{-1}\, \mathrm{Gpc}^{-3}\, h^{3}_{70}}

\title[ASAS-SN SNe Ia Rates]{Supernova Rates and Luminosity Functions from ASAS-SN I: \\2014--2017 Type Ia SNe and Their Subtypes}

\author[D.~D.~Desai et al.]{
D.~D.~Desai\orcidlink{0000-0002-2164-859X}$^{1}$\thanks{E-mail: dddesai@hawaii.edu},
C.~S.~Kochanek$^{2,3}$,
B.~J.~Shappee\orcidlink{0000-0003-4631-1149}$^{1}$,
T.~Jayasinghe\orcidlink{0000-0002-6244-477X}$^{4,\dagger}$,
K.~Z.~Stanek$^{2,3}$,
T.~W.-S.~Holoien\orcidlink{0000-0001-9206-3460}$^{5}$,
\newauthor
T.~A.~Thompson\orcidlink{0000-0003-2377-9574}$^{2,6,3}$,
C.~Ashall\orcidlink{0000-0002-5221-7557}$^{7}$,
J.~F.~Beacom\orcidlink{0000-0002-0005-2631}$^{6,2,3}$,
A.~Do\orcidlink{0000-0003-3429-7845}$^{1}$,
Subo~Dong\orcidlink{0000-0002-1027-0990}$^{8}$,
J.~L.~Prieto$^{9,10}$
\\
$^{1}$ Institute for Astronomy, University of Hawai`i at M\=anoa, 2680 Woodlawn Drive, Honolulu, HI 96822, USA \\
$^{2}$ Department of Astronomy, The Ohio State University, 140 West 18th Avenue, Columbus, OH 43210, USA\\
$^{3}$ Center for Cosmology and Astro-Particle Physics, The Ohio State University, 191 West Woodruff Avenue, Columbus, OH 43210, USA\\
$^{4}$Department of Astronomy, University of California, Berkeley, CA 94720\\
$^{5}$ Carnegie Observatories, 813 Santa Barbara Street, Pasadena, CA 91101, USA\\
$^{6}$ Department of Physics, The Ohio State University, 191 West Woodruff Avenue, Columbus, OH 43210, USA\\
$^{7}$ Department of Physics, Virginia Tech, Blacksburg, VA 24061, USA \\
$^{8}$ Kavli Institute for Astronomy and Astrophysics, Peking University, Yi He Yuan Road 5, Hai Dian District, Beijing 100871, China\\
$^{9}$ Núcleo de Astronomía de la Facultad de Ingeniería y Ciencias, Universidad Diego Portales, Av. Ejército 441, Santiago, Chile\\
$^{10}$ Millennium Institute of Astrophysics, Santiago, Chile\\
$^{\dagger}$ NASA Hubble Fellow
}

\date{Accepted XXX. Received YYY; in original form ZZZ}

\pubyear{2023}

\begin{document}
\label{firstpage}
\pagerange{\pageref{firstpage}--\pageref{lastpage}}
\maketitle

\begin{abstract}
We present the volumetric rates and luminosity functions (LFs) of Type Ia supernovae (SNe Ia) from the $V$-band All-Sky Automated Survey for Supernovae (ASAS-SN) catalogues spanning discovery dates from UTC~2014-01-26 to UTC~2017-12-29. Our standard sample consists of 404 SNe Ia with $m_{\mathrm{\textit{V},peak}} < 17\,\mathrm{mag}$ and Galactic latitude $|b|>15\degr$. Our results are both statistically more precise and systematically more robust than previous studies due to the large sample size and high spectroscopic completeness. We make completeness corrections based on both the apparent and absolute magnitudes by simulating the detection of SNe Ia in ASAS-SN light curves. We find a total volumetric rate for all subtypes of $R_{\mathrm{tot}} = 2.28^{+0.20}_{-0.20} \rateunits$ for $M_{\mathrm{\textit{V},peak}} < -16.5\,\mathrm{mag}$ ($R_{\mathrm{tot}} = 1.91^{+0.12}_{-0.12} \rateunits$ for $M_{\mathrm{\textit{V},peak}} < -17.5\,\mathrm{mag}$) at the median redshift of our sample, $z_{\mathrm{med}}=0.024$. This is in agreement ($1\sigma$) with the local volumetric rates found by previous studies. We also compile luminosity functions (LFs) for the entire sample as well as for subtypes of SNe Ia for the first time. The major subtypes with more than one SN include Ia-91bg, Ia-91T, Ia-CSM, and Ia-03fg with total rates of $R_{\mathrm{Ia-91bg}} = 1.4^{+0.5}_{-0.5} \times 10^{3}\,\mathrm{yr}^{-1}\, \mathrm{Gpc}^{-3}\, h^{3}_{70}$, $R_{\mathrm{Ia-91T}} = 8.5^{+1.6}_{-1.7} \times 10^{2}\,\mathrm{yr}^{-1}\, \mathrm{Gpc}^{-3}\, h^{3}_{70}$, $R_{\mathrm{Ia-CSM}} = 10^{+7}_{-7}\,\mathrm{yr}^{-1}\, \mathrm{Gpc}^{-3}\, h^{3}_{70}$, and $R_{\mathrm{Ia-03fg}} = 30^{+20}_{-20}\,\mathrm{yr}^{-1}\, \mathrm{Gpc}^{-3}\, h^{3}_{70}$, respectively. We estimate a mean host extinction of $E(V-r) \approx 0.2\,\mathrm{mag}$ based on the shift between our $V$ band and the Zwicky Transient Facility $r$-band LFs.
\end{abstract}

\begin{keywords}
methods: data analysis -- surveys -- supernovae: general
\end{keywords}

%%%%%%%%%%%%%%%%%%%%%%%%%%%%%%%%%%%%%%%%%%%%%%%%%%

%%%%%%%%%%%%%%%%% BODY OF PAPER %%%%%%%%%%%%%%%%%%

\section{Introduction} \label{sec:intro}
Type Ia supernovae (SNe Ia) are thermonuclear explosions of carbon/oxygen (C/O) white dwarfs \citep[WDs; e.g.,][]{hoyle60}. SNe Ia have been widely used as standardizable candles \citep[e.g.,][]{Pskovskii77,phillips93} to measure relative cosmological distances \citep[e.g.,][]{riess98,perlmutter99}. Although the progenitor systems of SNe Ia are uncertain, all likely scenarios include the interaction of a WD with another star \citep[for reviews see][]{maoz14,ruiter20}. The two basic models are the single-degenerate (SD) scenario, where the WD accretes from a non-degenerate companion star \citep[e.g.,][]{whelan73,nomoto82}, and the double-degenerate (DD) scenario, where both objects are WDs \citep[e.g.,][]{tutukov79,iben84,webbink84,Shen18}. Most observational studies disfavor the SD scenario for the majority of SNe \citep[e.g.,][]{nugent11,chomiuk12,shappee13,shappee18,tucker22b}.
However, the uncertain nature of the SNe Ia progenitor systems and explosion mechanisms remains a substantial problem for understanding potential systematic errors in using SNe Ia for cosmology \citep{Betoule14}.

It is also becoming apparent that SNe Ia are not a uniform class of objects; rather, there is a growing number of subtypes. These include the overluminous 91T-like SNe Ia \citep[e.g., ][]{Filippenko92a,Phillips92}; the subluminous 91bg-like SNe Ia \citep[e.g., ][]{Filippenko92b,Leibundgut93,Turatto96}; 03fg-like SNe Ia, which are brighter in the near-infrared and have long rise times \citep[e.g., ][]{Howell06,Hicken07,Hsiao20,Ashall21,Jiang21,Lu21}; 02cx-like SNe Ia, which show light curves that are both broad and faint \citep[e.g., ][]{Li03,Foley13}; the subluminous 02es-like SNe Ia, with the broad, slowly declining light curves seen in overluminous SNe Ia, yet lacking a prominent secondary maximum in the $i$ band as seen in subluminous SNe Ia \citep[e.g., ][]{Foley10,Ganeshalingam12}. There are still additional transitional subtypes \citep[for a review of subtypes, see][]{Taubenberger17}.

Different progenitor scenarios operate over a wide range of timescales, ranging from $\sim$100 Myr to the Hubble time. The rate as a function of the time span $\tau$ between a burst of star formation and the resulting SNe Ia is known as the delay-time distribution (DTD). Measurements of the DTD of SNe Ia \citep{maoz-mannucci12} can be used to constrain progenitor models and the convolution of the DTD with the cosmic star formation history (SFH) gives the redshift evolution of SN Ia rate. Most DD models predict that the DTD is a $\sim\tau^{-1}$ power-law \citep[e.g.][]{ruiter09,Horiuchi10,mennekens10} for longer delay times and observations of SN Ia rates are broadly consistent with such a model \citep[e.g.][]{scannapieco05,maoz12,graur13,graur14}. However, the SD scenario predicts a broad range of DTDs which fail to account for long delay times \citep[e.g.,][]{graur14}. 

The shorter delay times are attributed to 91T-like SNe Ia, which are predominantly found in late-type, star-forming galaxies \citep{Howell01,li11b} and are therefore likely associated with young stellar populations. The longer delay times are associated with 91bg-like SNe Ia, which are mostly found in massive early-type galaxies with star-formation rates below $\sim 10^{-9}\,\mathrm{M_{\odot}\,yr^{-1}}$ \citep{Howell01,Neill09,Gonzalez-Gaitanetal11}. Similar to their 91bg-like cousins, 02es-like events have a tendency to preferentially, but not exclusively, explode in massive, early-type host galaxies \citep{White15}. To better understand the diversity in the subtypes of SNe Ia and their contribution to the DTD, we must understand the rates of the SNe~Ia subtypes.

Large surveys have made it possible to obtain SN rate measurements and luminosity functions (LFs) to constrain the DTD and probe SNe Ia physics. Surveys such as the Lick Observatory Supernova Search \citep[LOSS;][]{li00}, the Palomar Transient Factory \citep[PTF;][]{rau09,law09}, the Sloan Digital Sky Survey (SDSS)-II Supernova Survey \citep{Frieman08}, the Panoramic Survey Telescope \& Rapid Response System \citep[Pan-STARRS;][]{flewelling20,chambers16}, the All-Sky Automated Survey for Supernovae \citep[ASAS-SN;][]{shappee14,kochanek17}, the Asteroid Terrestrial-impact Last Alert System \citep[ATLAS;][]{tonry18,smith20}, and the Zwicky Transient Facility \citep[ZTF;][]{masci19,bellm19} have discovered thousands of SNe. Hundreds of these SNe are spectroscopically classified \citep[e.g.,][]{Smartt15,Tucker22a}, especially the more local ones. 

From these and other samples, volumetric rates of normal Type Ia SNe have been measured over a wide range of redshifts. Local ($z<0.1$) rates were measured by \citet{cappellaro99} using 70 SNe Ia from heterogeneous sources, by \citet{li11b} using 274 SNe Ia from LOSS, by \citet{frohmaier19} using 90 SNe Ia from PTF, by \citet{perley20} using 875 SNe from the ZTF Bright Transient Survey (BTS) sample, and by \citet{sharon-kushnir22} using a lower redshift volume-limited subset of 298 SNe from the ZTF sample. The PTF and the ZTF BTS samples have the highest spectroscopic completeness ($93\%$) among the previous studies. They all agree at the $\sim1\sigma$ level. At higher redshifts, \citet{dilday10} measured SN Ia rates within $z<0.3$ using 270 spectroscopically classified SNe Ia from SDSS-II SN survey and \citet{perrett12} measured the rates up to $z\sim1.1$ using 286 spectroscopically classified SNe Ia from the SuperNova Legacy Survey (SNLS). Other studies include the Institute for Astronomy Deep Survey in the redshift range $0.1 < z < 1.05$ \citep[][]{Rodney10} and the Subaru Deep Field out to $z\sim2$ \citep[][]{graur11}. These observed SN rates have been used to compare different DTD models and show that a $\sim\tau^{-1}$ DTD successfully describes the observed rates \citep[e.g.,][]{Horiuchi10,maoz12,graur14}.

The creation of ASAS-SN was largely motivated by the incompleteness of the
local census of supernovae.  Prior to that, discoveries were dominated by amateurs,
favored large galaxies, and showed year-to-year fluctuations
that were too large to be random.  In \citet{holoien17a, holoien17b, holoien17c, holoien19}, ASAS-SN
cataloged the supernovae found in its first five years (2013--2017).
This sample includes 704 Type Ia SNe with 97\% spectroscopically classified. The limiting magnitude of ASAS-SN in the $V$ band ($\sim$17\,mag) made spectroscopic follow-up possible without the use of large telescopes, leading to high spectroscopic completeness. In addition to confirming the bias of the amateurs towards luminous hosts, \citet{holoien17a, holoien17b, holoien17c, holoien19} found that both amateur and other professional
surveys were strongly biased against finding SNe close to the centers of
galaxies. The median radial offset of the 2013-2017 ASAS-SN sample was $2.4\, \mathrm{kpc}$ compared to $5.7\, \mathrm{kpc}$ and $4.5\, \mathrm{kpc}$ for the amateur and other professional surveys, respectively. 

The first statistical SNe Ia study with ASAS-SN \citep{brown19} extended the finding by \citet{li11a} that the specific SNe Ia rate increases
for lower mass galaxies from $\sim 3$ decades in mass to $\sim 6$ ($6.3 \leq \log(M_*/M_{\odot}) \leq 12.3$).
\citet{brown19} found that the rate per unit stellar mass $M_*$ scales roughly as $M_*^{-1/2}$ for the \citet{Bell03}
stellar mass function ($M_*^{-1/3}$ for the \citealt{Baldry12} stellar mass function, see \citealt{Gandhi22}).
For the \citet{li11a} mass range, this could be explained by lower mass galaxies having younger stellar populations \citep{Kistler13,graur13}, but this solution does not work for even lower masses. Using Feedback In Realistic Environments (FIRE-2) cosmological zoom-in simulations, \citet{Gandhi22} found that including an SN Ia rate that increases with decreasing metallicity ($Z^{-0.5}$ to $Z^{-1}$) significantly improves agreement with observations. \citet{Johnson22} then used simple numerical calculations using mean star formation histories from the UniverseMachine \citep{Behroozi19}, a $\tau^{-1}$ SN Ia DTD \citep[e.g.,][]{maoz-mannucci12}, and the mass-metallicity relation for galaxies \citep[e.g.,][]{Tremonti04,Andrews13,Zahid11,Zahid14} and found that a $\sim Z^{-0.5}$ scaling is required.  Specifically, \citet{Johnson22} found that the combination of younger ages and lower metallicities for lower-mass galaxies can explain the scaling over the full mass range of \citet{brown19}. Both \citet{Gandhi22}  and  \citet{Johnson22}  proposed that a likely explanation for the SN rate-metallicity trend is the rapid rise in the binary fraction towards lower metallicity \citep{Badenes18,Moe19,Wyse20}.

In this study, we use Type Ia SNe from the ASAS-SN catalogues to measure the local volumetric rate and LF of Type Ia SNe, including, for the first time, LFs for several major spectroscopic subtypes. 
In Section~\ref{sec:data}, we describe the supernova sample and refit the data to update the peak apparent magnitudes in ASAS-SN. In Section~\ref{sec:rates_method}, we outline our approach to making completeness corrections using simulations and the method for calculating the rates. In Section~\ref{sec:results}, we present our local volumetric rates and LFs for Type Ia SNe and several spectroscopic subtypes and compare them to earlier results. We also provide the LFs corrected assuming various global values of the host-galaxy extinction. Finally, in Section~\ref{sec:summary}, we summarize the results and discuss future projects. Throughout this paper we adopt a flat Lambda cold dark matter cosmology with a Hubble constant $H_0 = 70 \,\mathrm{km\,s}^{-1}\,\mathrm{Mpc}^{-1}$ and a matter density $\Omega_{m,0} = 0.3$.

\section{The Supernova Sample} \label{sec:data}
\begin{figure}
	\includegraphics[width=\columnwidth]{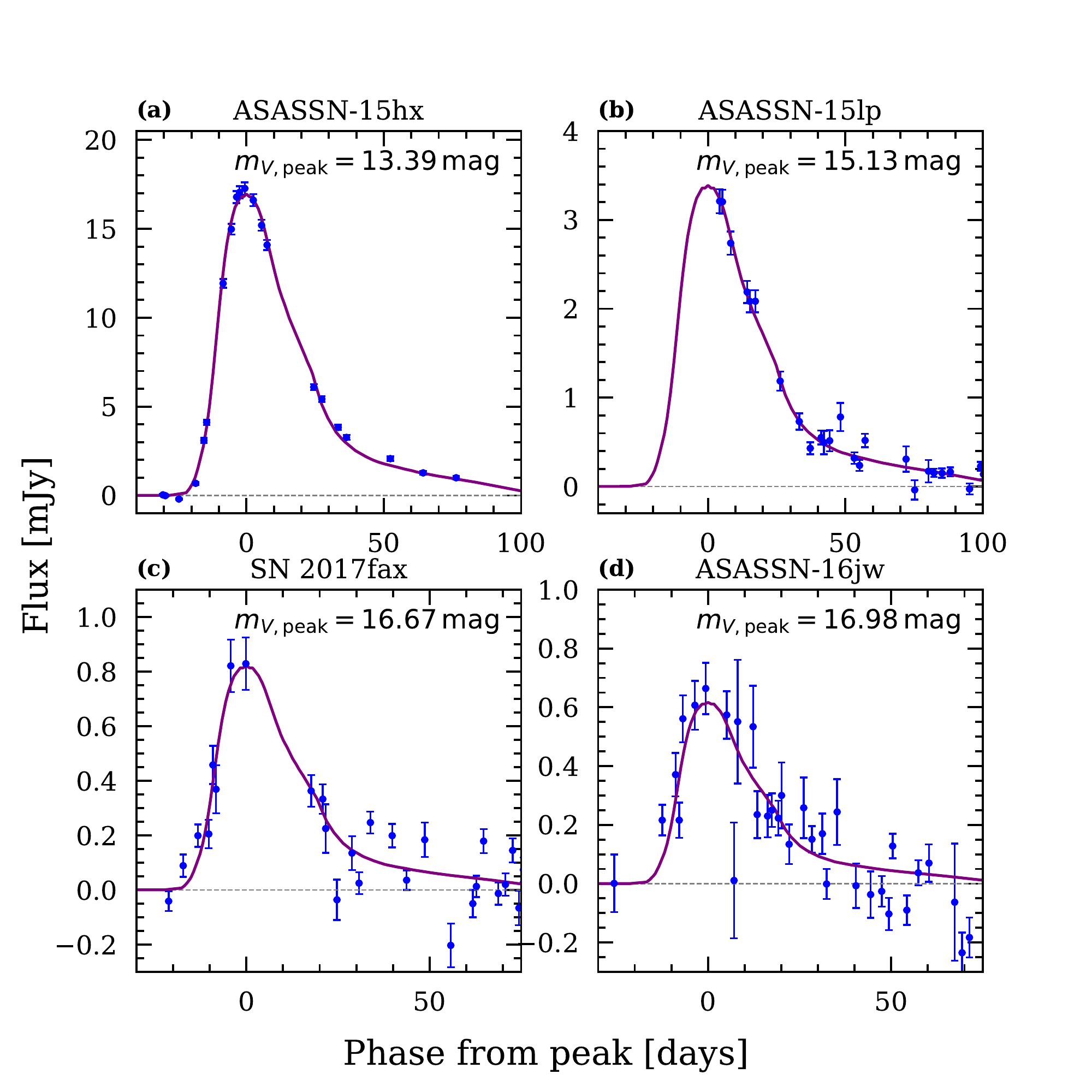}
    \caption{ASAS-SN $V$-band light curves (blue points) of four SNe (ASASSN-15hx, ASASSN-15lp, SN~2017fax, and ASASSN-16jw) spanning a range in peak magnitudes along with the best fits using the \citet[][]{nugent02} $V$-band SN Ia templates (purple line). The phase is given in the observed frame.}
    \label{fig:LC_fits}
\end{figure}

We use the Type Ia supernova sample of 704 SNe~Ia from the $V$-band ASAS-SN Bright Supernova Catalogues \citep[][]{holoien17a, holoien17b, holoien17c, holoien19} spanning discovery dates from UTC 2014-01-26 to UTC 2017-12-29. We perform our analysis on the 578 SNe~Ia discovered or recovered in ASAS-SN. SNe discovered in the $g$ band but never recovered in the $V$ band are excluded from the analysis. 

The ASAS-SN catalogues defined the peak magnitude of a supernova by taking the brighter value between the brightest point in the light curve and the peak of a parabolic fit to magnitudes. However, we found that this method systematically biases the peak magnitudes to be too bright since random fluctuations often make the brightest point brighter than the peak of a fit. This is primarily a problem for the faintest SNe in the catalogue and, unfortunately, most SNe are faint, with the median peak magnitude being 16.4\,mag.

To obtain more accurate values for the peak magnitudes, we refit all ASAS-SN light curves in flux instead of magnitude using the $V$-band SN Ia templates from \citet[][]{nugent02}. Using the templates, varying the stretch in time, time of peak, and peak magnitude, and using the fluxes instead of magnitudes leads to more robust fits for fainter SNe. The median peak magnitude of our new fits is $0.3\, \mathrm{mag}$ fainter than for the original approach. Unfortunately, the stretch is generally poorly constrained from the ASAS-SN light curves alone, as the fainter SNe are too noisy. In Table~\ref{tab:SNeIa}, we report the updated peak magnitudes for all SNe~Ia with ASAS-SN $V$-band light curves. We experimented with the Spectral Adaptive Light Curve Template \citep[SALT2;][]{Guy07} templates and found consistent estimates of the peak magnitudes.

\begin{table*}
\centering
\caption{The $V$-band Sample of Type Ia SNe Discovered or Recovered by ASAS-SN.}
\label{tab:SNeIa}
\resizebox{\textwidth}{!}{
\begin{tabular}{lccccccccccc}
\hline
\hline    
     SN Name & IAU Name & RA &   Dec. &  Redshift & Type & $m_{\mathrm{\textit{V},peak}}$ $^a$ & $M_{\mathrm{\textit{V},peak}}$ $^b$ & $s$ $^c$ & $t_{\mathrm{peak}}$ & Host Name  & Disc./Rec. $^d$ \\
     
             &   &  [J2000]      &  [J2000]  &  &      &  [mag]    &      [mag]       & & [JD] &   &\\
\hline
                  ASASSN-14ad &     --- & 12:40:11.10 &  $+$18:03:32.8 &  $0.0264$ &      Ia &      $16.15$ &  $-18.65$ & $1.04$ & $2456696.9$ &                   KUG 1237+183 &       D \\
                  ASASSN-14ar &     --- & 09:09:41.68 &  $+$37:36:07.6 &  $0.0230$ & Ia-91bg &      $16.41$ &  $-18.65$ & $0.96$ & $2456769.4$ &                         IC 527 &       D \\
                  ASASSN-14as &     --- & 12:57:34.11 &  $+$35:31:35.8 &  $0.0374$ &      Ia &      $17.13$ &  $-18.99$ & $0.80$ & $2456775.0$ &                 MGC +06-29-001 &       D \\
                  ASASSN-14ax &     --- & 17:10:00.68 &  $+$27:06:20.1 &  $0.0330$ &      Ia &      $16.99$ &  $-18.94$ & $1.07$ & $2456792.4$ &       SDSS J171000.69+270619.5 &       D \\
                  ASASSN-14ba &     --- & 10:21:31.72 &  $+$08:24:18.6 &  $0.0327$ &  Ia-91T &      $16.77$ &  $-19.08$ & $0.81$ & $2456798.1$ &       SDSS J102131.91+082419.8 &       D \\
                  ASASSN-14bb &     --- & 12:14:11.35 &  $+$38:39:40.8 &  $0.0230$ &      Ia &      $16.14$ &  $-18.91$ & $0.98$ & $2456801.0$ &        2MASX J12141125+3839400 &       D \\
                  ASASSN-14ay &     --- & 15:57:02.99 &  $+$37:24:56.4 &  $0.0309$ &      Ia &      $16.52$ &  $-19.20$ & $0.87$ & $2456796.4$ &        2MASX J15570268+3725001 &       D \\
                  ASASSN-14bd &     --- & 12:52:44.85 &  $+$26:28:13.1 &  $0.0214$ & Ia-91bg &      $16.93$ &  $-17.94$ & $0.84$ & $2456802.7$ &                         IC 831 &       D \\
                    iPTF14bdn &     --- & 13:30:44.88 &  $+$32:45:42.4 &  $0.0156$ &  Ia-91T &      $14.80$ &  $-19.38$ & $1.10$ & $2456824.8$ &                      UGC 08503 &       R \\
                  ASASSN-14bt &     --- & 10:19:19.74 &  $+$58:30:19.7 &  $0.0289$ &      Ia &      $16.48$ &  $-19.06$ & $0.91$ & $2456814.9$ &                       UGC 5566 &       D \\
                  ASASSN-14cb &     --- & 13:08:14.45 &  $+$62:02:02.5 &  $0.0336$ &      Ia &      $16.98$ &  $-18.90$ & $0.80$ & $2456823.2$ &         2MASXi J1308145+620200 &       D \\
                  ASASSN-14co &     --- & 15:57:29.75 &  $+$01:06:34.0 &  $0.0333$ &      Ia &      $16.95$ &  $-19.13$ & $1.10$ & $2456819.7$ &                   CGCG 023-005 &       D \\
                     LSQ14cnm &     --- & 16:05:24.50 &  $+$01:12:58.7 &  $0.0326$ &      Ia &      $16.89$ &  $-19.22$ & $1.10$ & $2456829.7$ &        2MASX J16052452+0113000 &       R \\
                  ASASSN-14cu &     --- & 12:47:02.61 &  $-$24:14:41.7 &  $0.0248$ &      Ia &      $17.36$ &  $-18.07$ & $0.80$ & $2456831.6$ &        2MASX J12470274-2414435 &       D \\
                       2014bv &  2014bv & 12:24:30.98 &  $+$75:32:08.6 &  $0.0056$ &      Ia &      $14.07$ &  $-17.94$ & $0.81$ & $2456840.9$ &                       NGC 4386 &       R \\
                  ASASSN-14db &     --- & 22:02:01.83 &  $-$70:02:27.9 &  $0.0375$ &      Ia &      $17.02$ &  $-19.15$ & $1.02$ & $2456827.7$ &                   ESO 075-G049 &       D \\
                  ASASSN-14dc &     --- & 02:18:38.06 &  $+$33:36:58.4 &  $0.0440$ &  Ia-CSM &      $15.93$ &  $-20.70$ & $1.10$ & $2456839.0$ &        2MASX J02183825+3336556 &       D \\
                  ASASSN-14dz &     --- & 15:05:54.52 &  $+$12:44:43.3 &  $0.0222$ &      Ia &      $16.03$ &  $-18.99$ & $0.99$ & $2456853.8$ &                        Mrk 842 &       D \\
                       2014by &  2014by & 14:27:49.00 &  $+$11:33:40.0 &  $0.0248$ &      Ia &      $16.35$ &  $-18.90$ & $0.82$ & $2456859.9$ &                      UGC 09267 &       R \\
\hline
\end{tabular}} \\
\begin{flushleft}
\noindent \textbf{Note:} This table includes 578 Type Ia SNe discovered or recovered in ASAS-SN from the catalogues \citep{holoien17a, holoien17b, holoien17c, holoien19} with updated peak apparent magnitudes $m_{\mathrm{\textit{V},peak}}$ and peak absolute magnitudes $M_{\mathrm{\textit{V},peak}}$. This table is available in its entirety in a machine-readable form in the online journal. A portion is shown here for guidance regarding its form and content. \\
$^a$ Peak apparent magnitudes reported here are the values after refitting the ASAS-SN light curves using the $V$-band SN Ia templates from \citet[][]{nugent02}. \\
$^b$ Peak absolute magnitudes are computed using $m_{\mathrm{\textit{V},peak}}$ and Eq.~\ref{eq:abs_mag}. \\
$^c$ Model stretch parameter related to $\Delta m_{15}(B)$ by Eq.\ref{eq:stretch_dm15}. $s=1.0$ corresponds to $\Delta m_{15}(B) = 1.05$.\\
$^d$ Indicates whether the SN was discovered by ASAS-SN (D) or independently recovered in ASAS-SN data (R).
\end{flushleft}
\end{table*}

Figure~\ref{fig:LC_fits} shows a few examples of the light-curve fits. Figure~\ref{fig:LC_fits}a and Figure~\ref{fig:LC_fits}d demonstrate the ability to fit one of the brightest (ASASSN-15hx) and one of the faintest (ASASSN-16jw) SNe in the sample. ASASSN-15lp, which only has a declining light curve (Figure~\ref{fig:LC_fits}b), would have had obvious issues for the peak magnitude based on either the brightest point or a parabolic fit near the peak. However, the templates encapsulate the shape of the decline and thus are able to predict the peak magnitude reasonably well. In the case of SN~2017fax (Figure~\ref{fig:LC_fits}c), with only two points near peak, a parabolic fit is poorly constrained, whereas the templates cover a larger time range leading to a better overall fit and peak magnitude.

We compute absolute magnitudes for all SNe using
\begin{equation}
    M_V = m_V - \mu (z) - A_{\mathrm{\textit{V},MW}} - K(z)
    \label{eq:abs_mag}
\end{equation}
where $m_V$ is the $V$-band apparent magnitude, $\mu (z)$ is the distance modulus as a function of redshift obtained using the Python package \texttt{astropy.cosmology} \citep[][]{astropy13,astropy18}, $A_{\mathrm{\textit{V},MW}}$ is the Milky Way extinction assuming a $7000\, \mathrm{K}$ source and $R_V = 3.1$ from Table 6 of \citet{sfd11}, and $K(z)$ is the $K$-correction \citep[][]{hogg02} as a function of redshift computed with SNooPy \citep[][]{burns11}, which uses the SN Ia templates from \citet{hsiao07}. We include no correction for the host-galaxy extinction here but explore its effects in Section~\ref{subsec:host_ext_LF}.

\begin{figure}
	\includegraphics[width=\columnwidth]{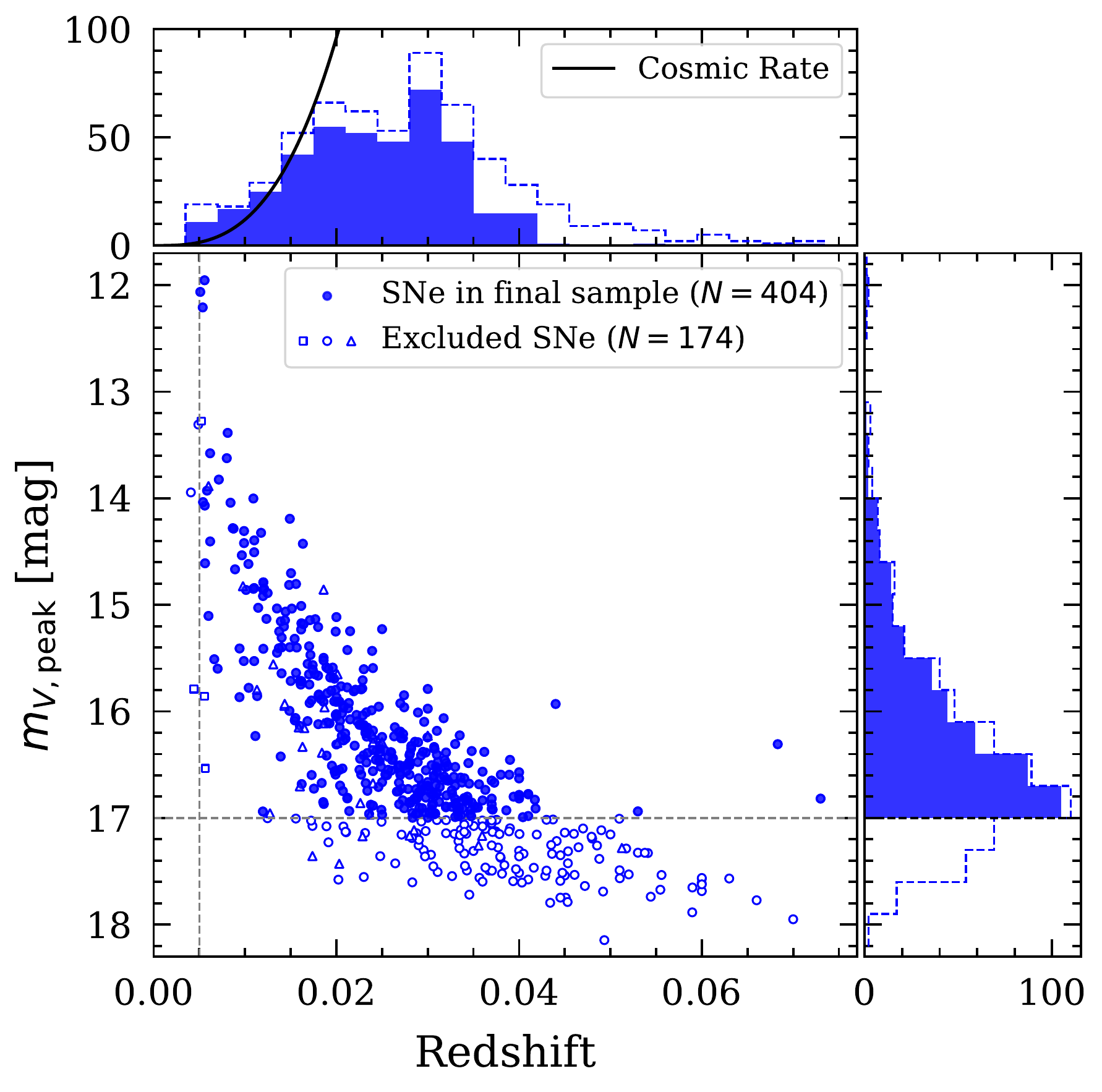}
    \caption{Peak apparent magnitude and redshift distribution of the 578 SNe Ia discovered or recovered by ASAS-SN. The filled circles and filled histograms show our standard sample of 404 SNe Ia. The open points are the SNe excluded due to the cuts on the peak absolute magnitude $M_{\mathrm{\textit{V},peak}}$ (open squares), Galactic latitude $b$ (open triangles), peak apparent magnitude $m_{\mathrm{\textit{V},peak}}$ and redshift $z$ (open circles). The gray dashed lines mark our choice of the limiting peak apparent magnitude at $m_{\mathrm{\textit{V},peak}}=17\, \mathrm{mag}$ and minimum redshift at $z_{\mathrm{min}}=0.005$. The cosmic rate (solid black line) shows that the sample is volume limited up to a redshift of $z\sim0.02$.}
    \label{fig:hists}
\end{figure}

We explore the statistics of the 578 SNe Ia discovered or recovered by ASAS-SN in the $V$ band \citep{holoien17a, holoien17b, holoien17c, holoien19} whose distribution in $m_{\mathrm{\textit{V},peak}}$ and redshift is shown in Figure~\ref{fig:hists}. 
The black line in the top histogram of Figure~\ref{fig:hists} displays the cosmic rate assuming a uniform distribution of SNe in comoving volume. It shows that the sub-sample with $M_{\mathrm{\textit{V},peak}} \lesssim 17.5\, \mathrm{mag}$ (which constitutes the majority of the sample) is volume-limited up to a redshift of $z\sim0.02$ and magnitude-limited beyond that.
This sample does not include the 13 SNe discovered in 2013 due to the high incompleteness of the survey and systematic errors during the early operations of ASAS-SN. 
We restrict our standard analysis to SNe more luminous than $M_{\mathrm{\textit{V},peak}}<-16.5\,\mathrm{mag}$, which reduces the sample to 574 SNe.
For our standard analysis we use a limiting Galactic latitude at $|b|>15\degr$, which leaves 541 SNe, and a limiting peak apparent magnitude at $m_{\mathrm{\textit{V},peak}} < 17\,\mathrm{mag}$, where our
completeness is $\sim 50\%$ (see Section~\ref{sec:rates_method}). These limits exclude the two
SNe~Iax from our standard sample; SN~2015H because we found a peak magnitude $> 17\,\mathrm{mag}$ and SN~2017gbb because it was not recovered by ASAS-SN. Finally, we restrict the redshift range to be from
$z_{\mathrm{min}}=0.005$, to eliminate systems where peculiar velocities can
significantly affect distance estimates, to $z_{\mathrm{max}}=0.08$, which
includes all systems. This leaves us with 404 SNe in our standard sample, where 
the reduced number is almost entirely due to the magnitude limit. 
We explore the consequences of varying these limits in Section~\ref{subsec:vol_rate}.

We update the subtype classification of three SNe in our standard sample to be consistent with the classification scheme of \citet{Taubenberger17}. ASASSN-15us was classified as Ia-06bt by \citet{holoien17b}, which falls under the broader class of Ia-02es. ASASSN-15hy and ASASSN-16ex were classified as Ia-07if and Ia-09dc, respectively, both of which fall under the broader class of Ia-03fg \citep{Ashall21}.

\section{Rate Computations}\label{sec:rates_method}
Magnitude-limited surveys like ASAS-SN must correct the observed sample for the probability of detecting the SNe. Incompleteness is driven by the survey cadence, seasonal gaps, survey magnitude limit, and observing conditions. As in previous studies, we use simulations to estimate the completeness corrections as a function of peak apparent and absolute magnitudes.

Based on all ASAS-SN observations of supernovae (discovered, recovered, and missed), we found that we could describe well the probability of a detection ($p$) in any given epoch as a simple function of the signal-to-noise ratio ($SNR$) of the observation
\begin{equation} \label{eq:det_model}
p = 
\left \{
    \begin{array}{ll}
        0,                                           & \text{for}\,\,SNR \leq 5 \\
        0.65 \left( 1 - \frac{12 - SNR}{7} \right),  & \text{for}\,\,5 \leq SNR \leq 12 \\
        0.65,                                        & \text{for}\,\,SNR \geq 12. \\
    \end{array}
\right.
\end{equation}
There was no apparent dependence on other variables. The saturation at $p = 0.65$ is due to a broad range of systematic problems associated with how the overall processing system tries to minimize the number of false positives.  However, this is a probability per observation, and with a $V$-band cadence of 3--4 days, the probability of actually missing a bright SN is very low. It is likely that Eq.~\ref{eq:det_model} is underestimating $p$ for bright SNe, but this also has no important consequences since bright SNe will have many detection trials ($n$) and the overall probability of detection $1-(1-p)^n$ converges rapidly to unity.  For example, if $n=5$, the probability of detection is $99.5\%$ for  $p=0.65$ versus $100.0\%$ for $p=0.95$.

Using Eq.~\ref{eq:det_model}, we perform injection recovery simulations on the ASAS-SN light curves. We obtain the ASAS-SN $V$-band light curves for random positions uniformly distributed on the sky with a mean density of four points per square degree. These light curves include the magnitudes, fluxes, and their uncertainties for each epoch and consequently include the survey cadence and seasonal gaps. Next, we inject simulated Type Ia SNe onto the light curves using the $V$-band templates from \citet[][]{nugent02} and ask which ones would be detected. We use the same magnitude and Galactic latitude limits as for the observed SN sample in Section~\ref{sec:data}. 
For bright SNe in bright galaxies, the presence of the galaxy will affect
the $SNR$, but the $SNR$ is so high that
neglecting the host does not matter for the detection probability. For
faint SNe, the noise is dominated by the sky, so the presence of the host
flux has little effect on the detection probability. 
Any systematic errors created by host galaxies will only become important once the statistical errors are smaller.

The time of peak $t_0$ for each simulated SN is drawn randomly from a uniform distribution over the time span covering the discovery dates of all SNe with a padding of 15 days on both ends. The redshift $z$ is drawn randomly assuming our standard cosmology and a constant comoving density with a maximum redshift of $z_{\mathrm{lim}}$ where the SN would have the limiting magnitude of $m_{\mathrm{\textit{V},peak}} = 17.0\, \mathrm{mag}$ given no extinction. The trial is kept if a uniform deviate is $< 1/(1+z)$ to account for the time dilation of the rates. The templates are stretched in time by the stretch parameter $s$ which is related to $\Delta m_{15}(B)$, the light curve decline rate parameter \citep[][]{phillips93}, by
\begin{equation} \label{eq:stretch_dm15}
    s = -0.397 \Delta m_{15}(B)^3 + 1.767 \Delta m_{15}(B)^2 - 3.034 \Delta m_{15}(B) + 2.698.
\end{equation}
This relation is obtained by fitting a third-order polynomial to the stretch factor as a function of $\Delta m_{15}(B)$ using the $B$-band templates from \citet{nugent02}, where $s=1.0$ corresponds to $\Delta m_{15}(B) = 1.05$. The $\Delta m_{15}(B)$ for each simulated SN is given by the $\Delta m_{15}(B)$ -- $M_{\mathrm{\textit{V},peak}}$ relation of
\begin{equation}\label{dm15_Mpeak}
    \Delta m_{15}(B) = 1.1 + 0.290\log \left[ 1 + \frac{M_{\mathrm{\textit{V},peak}} + 19.389 - 5 \log h_{70}}{0.096} \right]
\end{equation}
from \citet{garnavich04}, where $H_0 = 70\,\mathrm{km\,s^{-1}\,Mpc^{-1}}\,h_{70}$. Note that $\Delta m_{15}(B)$ is only an intermediate parameter to go from $M_{\mathrm{\textit{V},peak}}$ to the stretch $s$ that is applied to the $V$-band templates used to fit the light curves. The templates are further stretched in time by a factor of $(1+z)$ to account for the cosmological time dilation.

To account for the measurement uncertainty for the simulated SN light curve, we approximate the noise by adding a Gaussian deviate of dispersion
\begin{equation}
    \sigma_F = \sqrt{\frac{F_0 F}{g}},
\end{equation}
where $F$ is the flux for a given simulated point, $g$ is the gain for the camera used for observation at that epoch, and $F_0 = Z_0 10^{-m_0/2.5}$, where $Z_0$ and $m_0$ are the flux and magnitude zero-points, respectively.

\begin{figure}
	\includegraphics[width=\columnwidth]{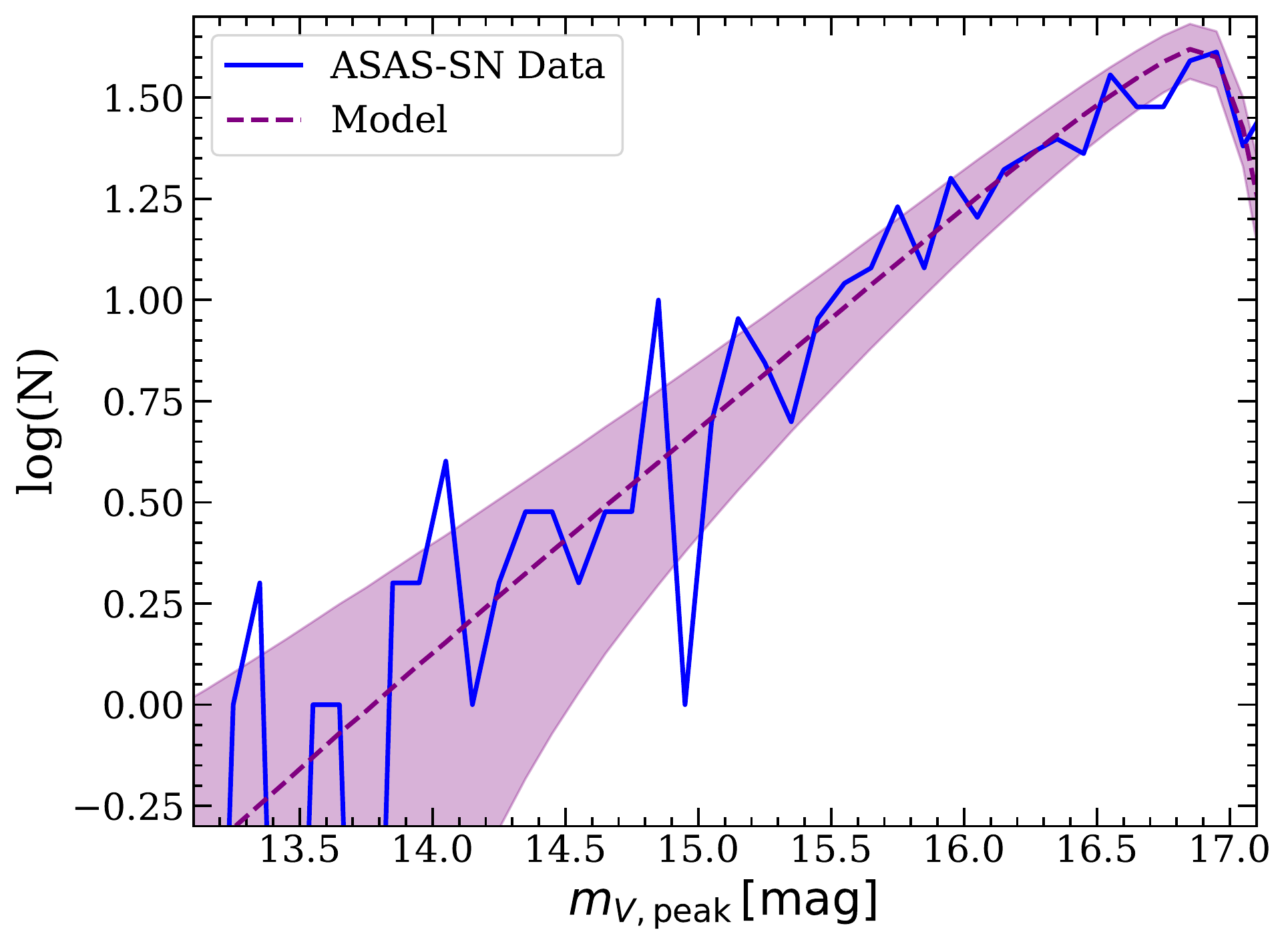}
    \caption{Differential distribution of peak apparent magnitudes $m_{\mathrm{\textit{V},peak}}$ for our observed SNe Ia sample (solid blue line) compared with the prediction from our detection model (dashed purple line). The model is normalized to have the same total number of SNe with $m_{\mathrm{\textit{V},peak}} < 17.0\, \mathrm{mag}$. The shaded region shows the $1\sigma$ Poisson uncertainty given the expected number of SNe.}
    \label{fig:V_dist}
\end{figure}

We carry out 100 random trials for each of the $N_{\mathrm{LC}} = 164{,}191$ random light curves, for a total of $M = 100\,N_{\mathrm{LC}}$ trial SNe. This is done for each peak luminosity over the range $-21\, \mathrm{mag} \leq M_{\mathrm{\textit{V},peak}} \leq -16.5\, \mathrm{mag}$ in intervals of $0.5\,\mathrm{mag}$. For each of these simulated SN, we use the $K$-correction and distance modulus appropriate to the random redshift, add the Galactic extinction correction associated with the light curve coordinates, and then add the flux of the SN to the random ASAS-SN light curve. The trial is logged as a detection if at least one epoch of the simulated SN satisfies the $SNR$ criterion in Eq.~\ref{eq:det_model}. 

The result from applying the detection model of Eq.~\ref{eq:det_model} to our simulated sample is shown as the dashed purple line in Figure~\ref{fig:V_dist} and it predicts the observed magnitude distribution of the sample very well despite having made no use of this information. The model shown here is normalized to the total number of SNe in the observed sample. The simulated sample of SNe not only reproduces the bright end of the observed distribution, but also well models the turnover at the faint end ($m_{\mathrm{\textit{V},peak}} \gtrsim 16.5\,\mathrm{mag}$).

\begin{figure}
	\includegraphics[width=\columnwidth]{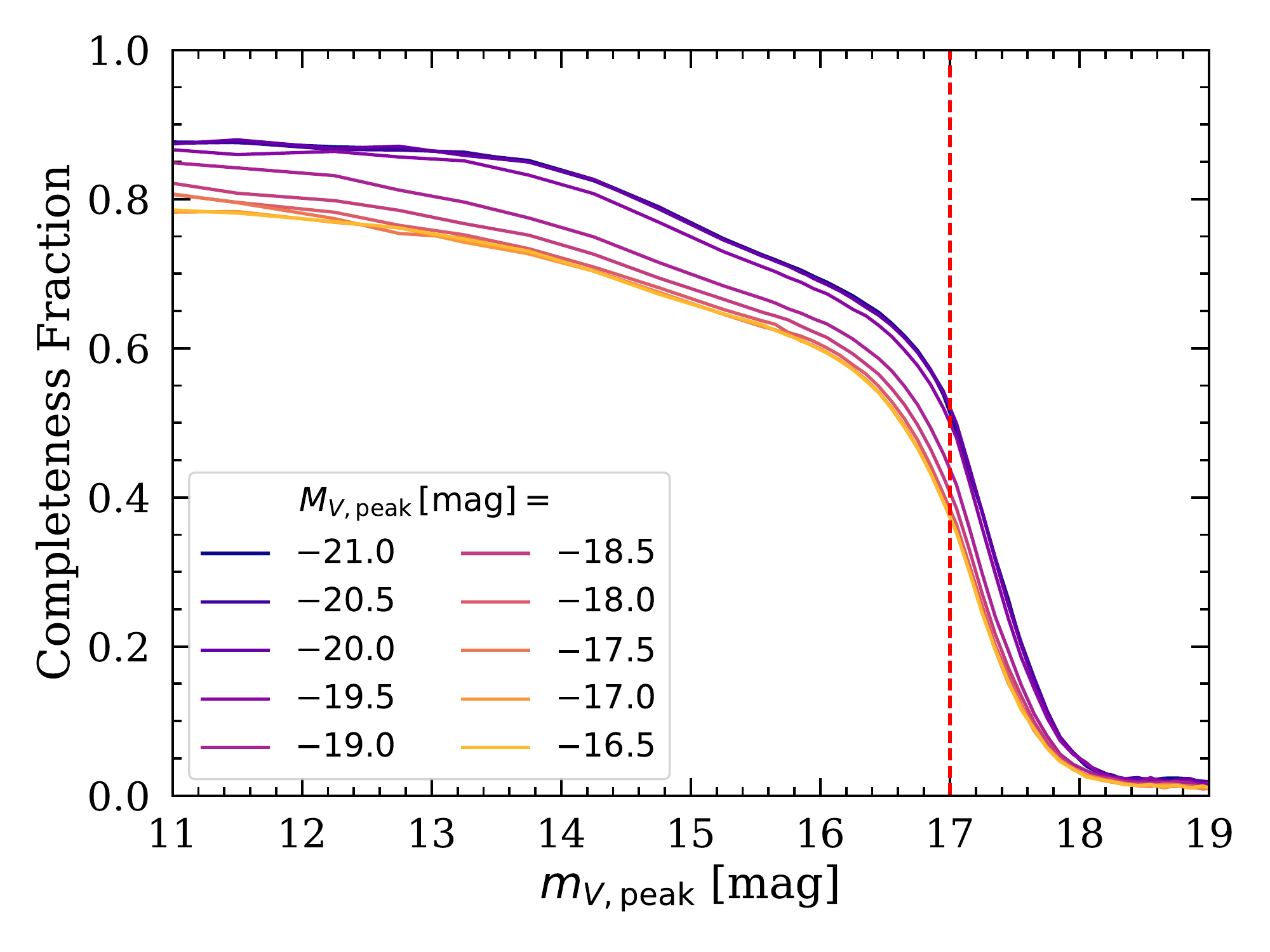}
    \caption{Completeness as a function of peak apparent magnitude $m_{\mathrm{\textit{V},peak}}$ with the peak absolute magnitudes $M_{\mathrm{\textit{V},peak}}$ shown in different colours. The vertical dashed red line marks our standard choice for the limiting magnitude, $m_{\mathrm{\textit{V},lim}} = 17\,\mathrm{mag}$, where the completeness is $\sim50\%$. More luminous SNe are luminous for a longer time, leading to higher completeness.}
    \label{fig:comp_vs_mag}
\end{figure}

Next, we compute the completeness as a function of peak absolute and apparent magnitudes. If a total of $N$ out of $M$ trials are detections, then the completeness is $F_1=N/M$. Figure~\ref{fig:comp_vs_mag} shows the completeness as a function of apparent magnitude for the different absolute magnitudes after binning the trials by their apparent magnitude. The completeness flattens out at the bright end at $\sim 80$ -- $85 \%$, limited by the seasonal gaps. The completeness slowly declines but then begins to drop rapidly from $\sim60\%$ at $m_{\mathrm{\textit{V},peak}} = 16\,\mathrm{mag}$ to $\sim0\%$ at $m_{\mathrm{\textit{V},peak}} = 18\,\mathrm{mag}$. The completeness is also lower for less luminous SNe because they spend less time near their peak luminosity. We choose a limit of $m_{\mathrm{\textit{V},lim}} = 17\,\mathrm{mag}$ for our standard analysis since that is where the completeness is $\sim50\%$. 

Since we adjust $z_{\mathrm{lim}}$ with the peak absolute magnitude to avoid wasting trials, we need to correct the completeness to a common volume for all SNe.  Our choice of $z_{\mathrm{max}} = 0.08$ defines the maximum redshift. Thus given the comoving volume $V(z)$, there is a second completeness factor of $F_2(M_{\mathrm{\textit{V},peak}}) = V(z_{\mathrm{lim}}(M_{\mathrm{\textit{V},peak}})) / V(z_{\mathrm{max}})$ to correct for the differences in volume between $z_{\mathrm{lim}}$ and $z_{\mathrm{max}}$. The final statistical weight for the $i^{\mathrm{th}}$ observed SN is $w_i = \left(F_{1,i}\, F_{2,i}\right)^{-1}$.

Given the statistical weights, the volumetric rate $R$ for SNe Ia is calculated by summing $N$ SNe within a time span $\Delta t$ and a fixed comoving volume $V$. Each SN is weighted by the factor $w_i$ that accounts for the incompleteness given its peak apparent and absolute magnitudes. The volumetric SN rate is then
\begin{equation}
    R = \frac{\sum_{i=1}^{N} w_i}{V\, \Delta t \, \left(1-\sin{b_{\mathrm{lim}}}\right)} ,
    \label{eq:rate}
\end{equation}
where $\Delta t = 4.0\, \mathrm{yr}$ is the time span between UTC 2014-01-01 and UTC 2017-12-31, 
$V = \frac{4}{3}\pi \left( d_{\mathrm{max}}^3 - d_{\mathrm{min}}^3 \right)$ is the total comoving volume corresponding to the maximum and minimum redshift limits, and $(1-\sin{b_\mathrm{lim}})$ corrects for our Galactic latitude limit. Including the lower redshift limit of $z_{\mathrm{min}}=0.005$ changes the volume by only $\sim0.03\%$. The effects of time dilation are already included in the computation of the weights. Therefore, the rate $R$, as given by Eq.~\ref{eq:rate}, provides an estimate for the number of SNe that occur per unit comoving volume and time.

We estimate the total volumetric rate using Eq.~\ref{eq:rate}. We also compute an LF by splitting the sample into absolute magnitude bins of width 0.5\,mag and use Eq.~\ref{eq:rate} to calculate the rate in each bin divided by the bin size to obtain the rate per magnitude. Because of the sample size and high spectroscopic completeness, we are also able to compute LFs for some of the major subtypes of SNe Ia. The subtypes that have more than one object include the overluminous Ia-91T ($N=30$), the subluminous Ia-91bg ($N=9$), the extremely bright Ia-CSM ($N=3$) and Ia-03fg ($N=2$) classes.

The statistical errors on all rates are estimated using bootstrapping. We first randomly draw the sample size as a Poisson deviate with an expected number equal to the observed sample ($N=404$ for our standard sample). Then we randomly draw that number of SNe from the observed sample with replacement. This also approximates the error in the completeness corrections coming from the SN weights through the random selection of SNe. The standard error on the rate is then given by the $16^{\mathrm{th}}$ and the $84^{\mathrm{th}}$ percentiles of the bootstrapped rate distribution. Errors for the bins in the LFs with more than one object are obtained in the same manner as the total sample. Errors for the bins with only one object are estimated using only the Poisson uncertainties.

We did not explicitly correct for the five SNe in the $V$-band catalogues without spectroscopic classification that could have been selected in our final sample. Given the relative numbers of SNe Ia and core-collapse SNe in \citet{holoien19}, we would expect $\sim3$ of these to be SNe Ia. Crudely, this implies an underestimate of the rates by $\sim3/404$, or less than $1\%$, which is much smaller than the statistical or other systematic uncertainties.

\section{Results \& Discussion}\label{sec:results}
\subsection{Volumetric SN Ia Rates} \label{subsec:vol_rate}
\begin{figure*}
	\includegraphics[width=\textwidth]{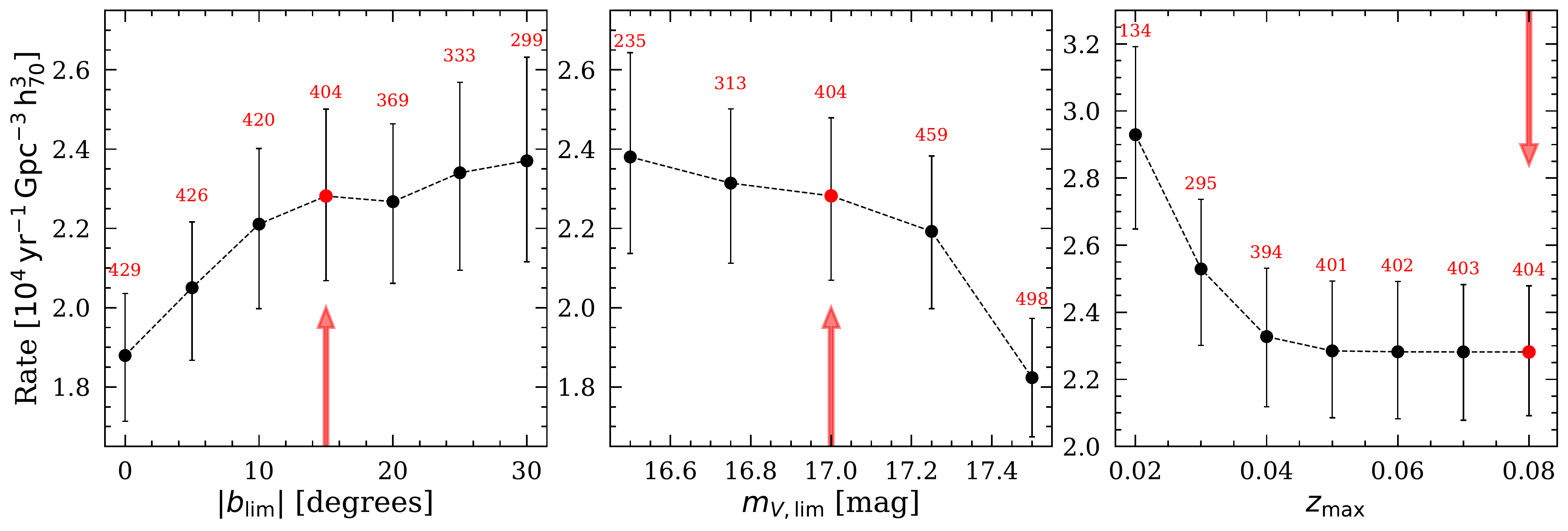}
    \caption{\textit{Left}: Rate as a function of the Galactic latitude cut. \textit{Middle}: Rate as a function of the limiting apparent magnitude. \textit{Right}: Rate as a function of maximum redshift cut. In all three panels, the numbers in red are the number of SNe in the samples. The red arrows point to the values used for our standard sample.}
    \label{fig:rate_vs_b_mlim_zmax}
\end{figure*}

Figure~\ref{fig:rate_vs_b_mlim_zmax} shows how our total rate estimates depend on the choice of the minimum Galactic latitude, limiting apparent magnitude, and maximum redshift. A fainter limiting magnitude or working closer to the
Galactic plane includes more SNe in the sample, thereby reducing the statistical errors, but
requires larger completeness corrections, which increases the systematic
uncertainties. The total volumetric rate is roughly constant for a latitude cut of $|b_\mathrm{lim}| > 15\degr$ and for a limiting magnitude of $m_{\mathrm{\textit{V},lim}} < 17\,\mathrm{mag}$, indicating that the completeness corrections are performing as expected given the uncertainties. As a compromise between the number of SNe and systematic uncertainty, we choose our standard values of $|b_\mathrm{lim}| = 15\degr$ and $m_{\mathrm{\textit{V},lim}} = 17\,\mathrm{mag}$. The total rate approaches a constant value with larger volume as $z_{\mathrm{max}}$ increases. There are no more SNe in our sample beyond $z=0.08$ so the rate is constant for higher redshifts. We choose $z_{\mathrm{max}} = 0.08$ as our standard limit.

Our SNe Ia sample has a median redshift of $z_{\mathrm{med}}=0.024$ and the total rate calculated using the method described in Section~\ref{sec:rates_method} is 
\begin{equation}
    R_{\mathrm{tot}} = 2.28^{+0.20}_{-0.20} \rateunits.
\end{equation}
The fractional uncertainty ($\sigma_R/R$) of 9\% is significantly more than the Poisson uncertainties from having 404 sources (5\%) in part because the weight factors $w_i$ are not uniform.  For example, if a small fraction of the sources have very high weights, the statistical uncertainties are ultimately controlled by the Poisson variations in the numbers of these highly weighted sources rather than the fluctuations in the larger number of overall sources.  In our case, this is driven by the minimum luminosity $M_{\mathrm{\textit{V},peak}}$ used to define the sample.  While the weight factor
depends weakly on absolute magnitude (Figure~\ref{fig:comp_vs_mag}), the volume correction basically scales as $\log w_i \propto -0.6\,M_{\mathrm{\textit{V},peak}}$,
which varies by a factor of $500$ between $M_{\mathrm{\textit{V},peak}} = -16.5\,\mathrm{mag}$ and $-21\,\mathrm{mag}$.
As shown in Table~\ref{tab:rates_vs_lowMcut}, increasing the minimum luminosity limit by a magnitude excludes few SNe from the sample. The total rate decreases as a result, since a smaller luminosity range is considered. But, excluding the lower luminosity SNe leads to uncertainties that approach the Poisson statistics limit.

\begin{table}
\centering
\caption{Volumetric SNe Ia Rates for Varying Minimum Luminosities.}
\label{tab:rates_vs_lowMcut}
% \resizebox{\columnwidth}{!}{
\renewcommand{\arraystretch}{1.5}
\begin{tabular}{cccc}
\hline
\hline
    $M_{\mathrm{\textit{V},peak}}$ & $N_{\mathrm{Ia}}$ & Rate & $\sigma_R/R$\\
    
    [mag] & & $[10^{4}\, \mathrm{yr}^{-1}\, \mathrm{Gpc}^{-3}\, h^{3}_{70}]$ & \\
\hline
    $[-16.5, -21.5]$ &  404   &  $2.28^{+0.20}_{-0.20}$ &  $9\%$ \\
    $[-17.0, -21.5]$ &  402   &  $2.09^{+0.15}_{-0.15}$ &  $7\%$ \\
    $[-17.5, -21.5]$ &  397   &  $1.91^{+0.12}_{-0.12}$ &  $6\%$ \\
\hline
\end{tabular}%}
\end{table} 

\begin{figure*}
	\includegraphics[width=\textwidth]{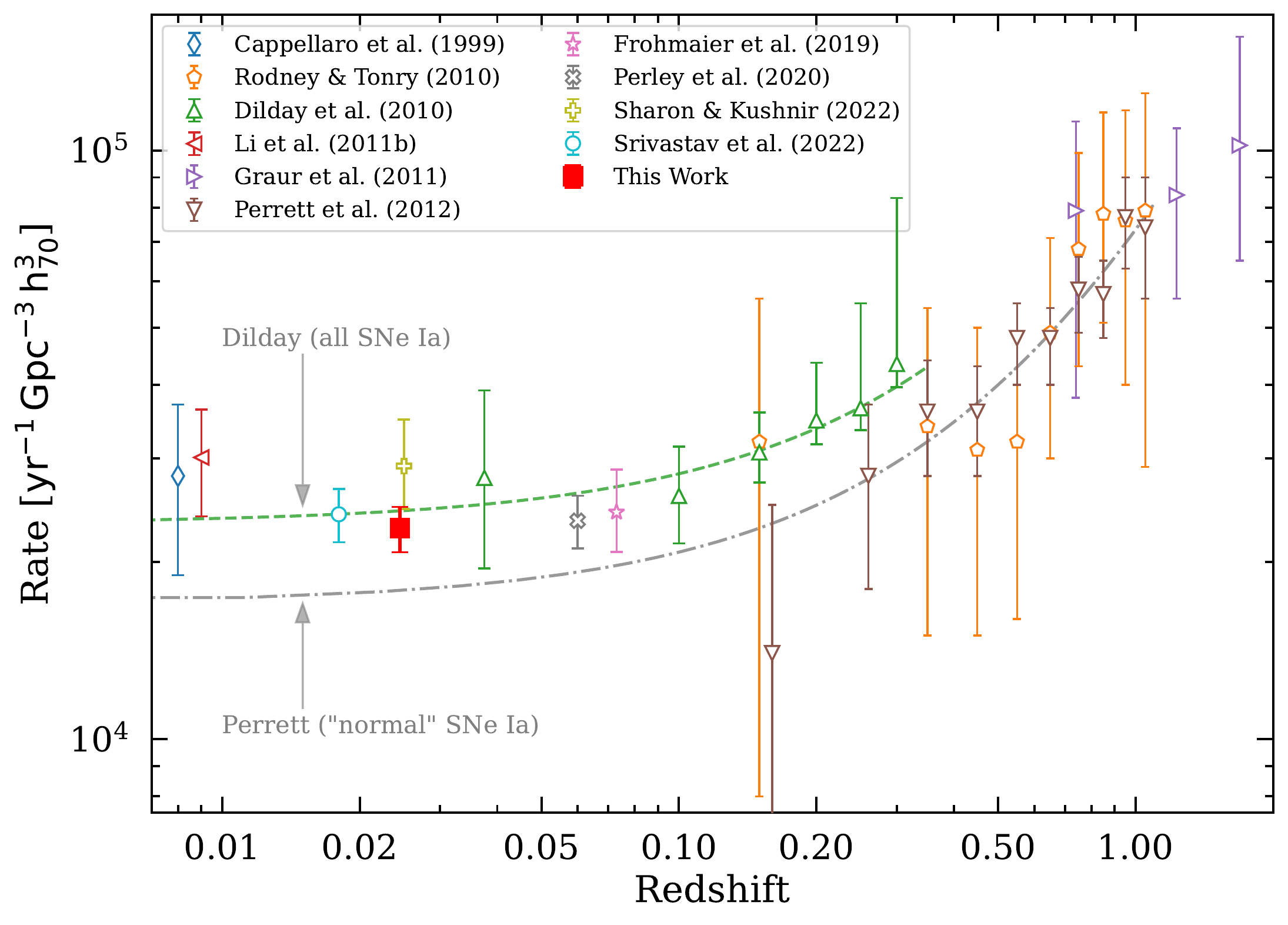}
    \caption{Volumetric SN Ia rate as a function of redshift $z$. Our rate is the filled red square at the median redshift of our sample, with other results as listed in the figure are shown as the open symbols. The dashed and dotted-dashed lines show a power-law fits of the form $(1+z)^{\alpha}$ from \citet[$\alpha = 2.04^{+0.90}_{-0.89}$;][]{dilday10} and \citet[$\alpha = 2.11 \pm 0.28$;][]{perrett12}, respectively. \citet{dilday10} includes all subtypes of SNe~Ia, whereas \citet{perrett12} only includes `normal' SNe~Ia. All displayed uncertainties include statistical and systematic.}
    \label{fig:rate_vs_z}
\end{figure*}

After converting the rates from various studies to a consistent value of $H_0 = 70 \,\mathrm{km\,s}^{-1}\,\mathrm{Mpc}^{-1}$, we first compare our total volumetric rate to other studies at low redshift ($z < 0.1$). The rates at low redshift are summarized in Table~\ref{tab:rates_vs_z} and are shown in Figure~\ref{fig:rate_vs_z}. Rates from both \citet{cappellaro99} and \citet{li11b} use a galaxy-targeted sample, which introduces large systematics in the observed sample. In particular, there was a bias towards more luminous galaxies, thus biasing the SN sample due to the correlation of SNe Ia and host properties \citep[e.g.,][]{Sullivan10}. However, due to their large uncertainties, these rate measurements agree with our total rate within $1\sigma$.

Untargeted surveys provide a means of reducing the galaxy bias, although there is still the question of the radial SN distribution. \citet{holoien17a,holoien17b,holoien17c,holoien19} showed that between 2014 and 2017, amateurs and surveys other than ASAS-SN were less effective in discovering SNe close to the centers of their hosts. Nonetheless, untargeted surveys have fewer systematics and show an improvement over targeted samples. 

Using the 270 spectroscopically confirmed SNe Ia from the SDSS-II Supernova Survey \citep{Frieman08}, \citet{dilday10} computed the rates out to a redshift of $z \sim 0.3$. Their lowest redshift bin ($0.025 \leq z \leq 0.050$) contains only four SNe, while their sample with $z<0.12$ contains 37 SNe. Due to a small sample size, their statistical uncertainties dominate, making their rates consistent with our total rate. \citet{frohmaier19} used a larger sample of 90 spectroscopically confirmed SNe Ia in the redshift range $z < 0.09$ from the untargeted PTF survey. The increase in the low-redshift sample size over the SDSS sample significantly reduces their statistical uncertainties. The PTF rate is consistent with the SDSS rate, and agrees with our total rate within $1\sigma$.

The largest untargeted sample is the ZTF BTS SNe Ia sample, going out to a redshift of $z \sim 0.1$. \citet{perley20} were able to benefit from the faint magnitude limit ($m \approx 18.5\,\mathrm{mag}$) of ZTF BTS to build a larger sample of 875 SNe. However, there are likely larger systematics since they used an average completeness correction rather than weighting individual SNe. Nonetheless, the total ZTF BTS rate is consistent with our total rate.

\citet{Srivastav22} computed the volumetric rate of SNe~Ia within a redshift of $z < 0.024$ using the ATLAS local volume survey and their estimate is consistent with ours. With 269 classified SNe~Ia in their sample, they perform an assessment of recovery of their simulated light curves given the history of ATLAS observations and quality metrics. They also compute a slightly higher rate of $(2.83 \pm 0.29) \times 10^{4}\,\mathrm{yr}^{-1}\, \mathrm{Gpc}^{-3}\, h^{3}_{70}$ including the spectroscopically unclassified SNe~Ia by assuming they have the same relative fractions of SN subtypes as the classified sample.

Using a subset of 298 SNe Ia within a redshift range $0.01 \leq z \leq 0.04$ from the ZTF BTS sample,
\citet{sharon-kushnir22} computed the volumetric rate after correcting for host-galaxy extinction. They obtained the intrinsic luminosities, corrected for host-galaxy extinction, based on the colour stretch $s_{gr}$ calibrated from the Carnegie Supernova Project SNe Ia sample \citep{Contreras10,Stritzinger11,Krisciunas17,Burns18,Ashall20}. Since they corrected for the host-galaxy extinction, the rate from \citet{sharon-kushnir22} is higher than our rate at a similar median redshift. We discuss host extinction in Section~\ref{subsec:host_ext_LF}.

\begin{table}
\centering
\caption{Local Volumetric SNe Ia Rates.}
\label{tab:rates_vs_z}
\resizebox{\columnwidth}{!}{
\renewcommand{\arraystretch}{1.5}
\begin{tabular}{cccl}
\hline
\hline
    $z$ & $N_{\mathrm{Ia}}$ & Rate & Reference  \\

     & & $[10^{4}\, \mathrm{yr}^{-1}\, \mathrm{Gpc}^{-3}\, h^{3}_{70}]$ & \\
\hline
    $\sim$0 &  70   &  $2.8 \pm 0.9$         & \citet{cappellaro99} \\
    0.025 -- 0.050    &  4   &  $2.78^{\,+1.13}_{\,-0.83}$  & \citet{dilday10} \\
    < 0.12    &  37   &  $2.35^{\,+0.47}_{\,-0.39}$  & \citet{dilday10} \\
    $\sim$0 &  274  &  $3.01 \pm 0.62$         & \citet{li11b} \\
    0.015 -- 0.090   &  90   &  $2.43^{\,+0.44}_{\,-0.35}$         & \citet{frohmaier19} \\
    < 0.1    &  875  &  $2.35 \pm 0.24$         & \citet{perley20} \\
    0.01 -- 0.04   &  298  &  $2.91^{\,+0.58}_{\,-0.45}$  & \citet{sharon-kushnir22} \\
    < 0.024  &  269  & $2.41 \pm 0.25$  & \citet{Srivastav22} \\
    \textbf{0.005 -- 0.08} &  \textbf{404}  &  \boldmath $2.28^{\,+0.20}_{\, -0.20}$ \unboldmath  & \textbf{This Work} \\
\hline
\end{tabular}}
\end{table} 

\begin{figure*}
	\includegraphics[width=\textwidth]{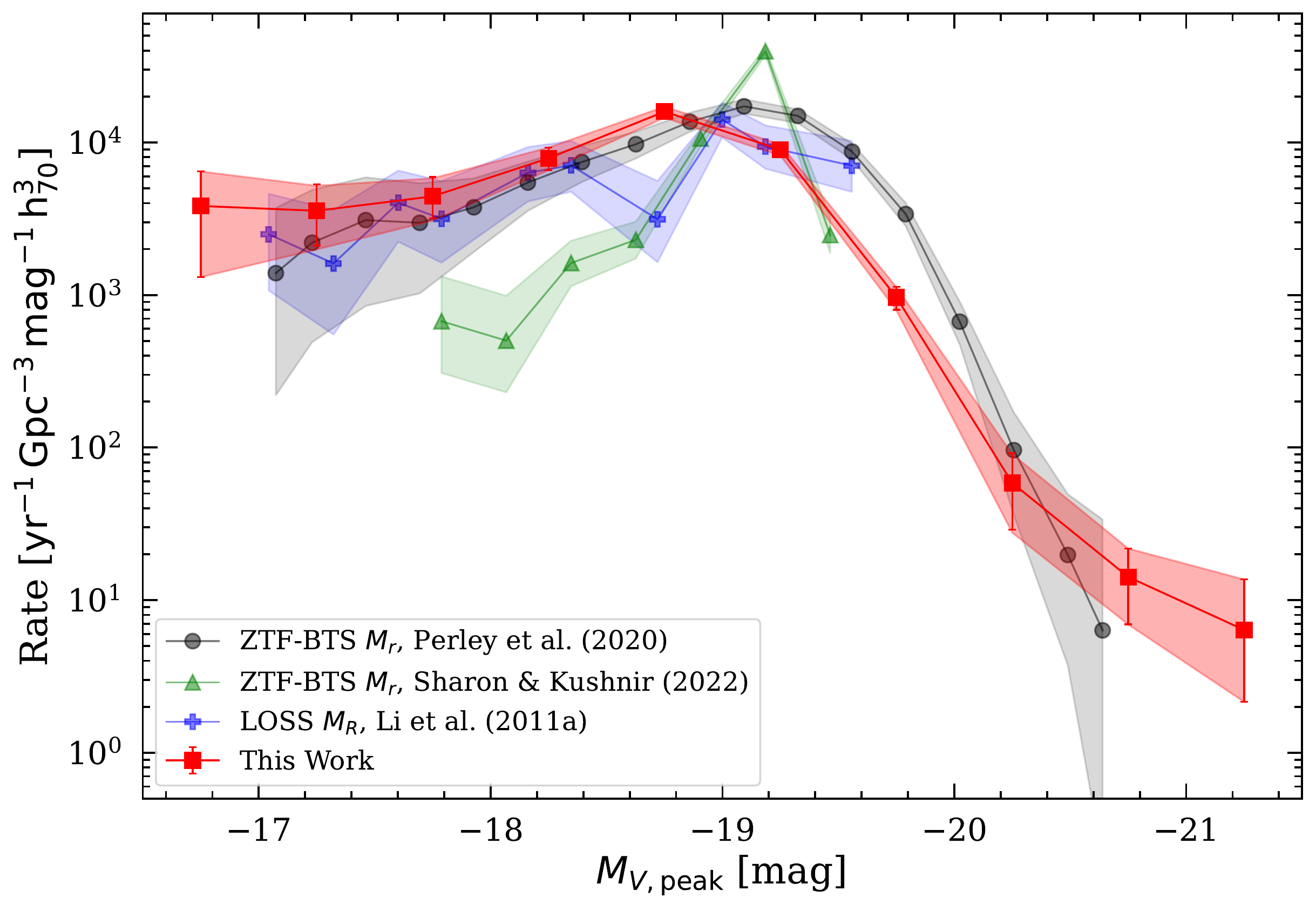}
    \caption{Luminosity functions of Type Ia SNe. Each magnitude bin has a width of $0.5\,\mathrm{mag}$ for the ASAS-SN data. $R$- and $r$-bands of \citet[][]{perley20}, \citet[][]{sharon-kushnir22}, and \citet[][]{li11a} are corrected to the $V$-band using Eq.~\ref{eq:r_to_v}. The LOSS LF is scaled to match the total ZTF BTS rate and shifted so their maxima coincide. All others are observed LFs, but the LF from \citet[][]{sharon-kushnir22} is an intrinsic LF, corrected for host-galaxy extinction. The two most luminous bins only contain SNe Ia-CSM.}
    \label{fig:rate_vs_M}
\end{figure*}

Figure~\ref{fig:rate_vs_z} also shows the rate as a function of redshift. Using SDSS data, \citet{dilday10} computed the rates in a broad range of redshifts, $0.025 \leq z \leq 0.325$, and fitted the redshift evolution with a power law of the form $(1+z)^{\alpha}$, with best-fit $\alpha = 2.0^{+0.9}_{-0.9}$ (dashed line in Figure~\ref{fig:rate_vs_z}) where the normalization is set by the rate at $z=0$. \citet{perrett12} measured the rates within a redshift range $0.1 \leq z \leq 1.1$ using SNe Ia from the SNLS. Their evolution of the rate with redshift is similarly fitted with a $(1+z)^{\alpha}$ power law with $\alpha = 2.1 \pm 0.3$ (dotted dashed line in Figure~\ref{fig:rate_vs_z}). However, their extrapolated rate at $z=0$ is lower than that of \citet{dilday10} because \citet{perrett12} did not include SNe that are subluminous Ia-91bg, super-Chandrasekhar, or extremely rare events. They limited their sample to `normal' SNe~Ia for better modelling. The result is a reduction in the overall normalization of the redshift evolution by $\sim15$ -- $20\%$. Our rate measurement, which includes all subtypes, is consistent with the fit from \citet{dilday10} and not with that from \citet{perrett12}. For both \citet{dilday10} and \citet{perrett12}, the DTD fit is consistent with $\propto \tau^{-1}$. The redshift evolution of rate from \citet{Rodney10} and \citet{graur11} agrees with that of \citet{perrett12} at higher redshifts but both have significantly larger uncertainties. Another approach, as shown in \citet{Horiuchi10}, is to use a star formation rate (SFR) to constrain the exponent of the DTD. However, this method requires an assumption for the SFR.

\subsection{Luminosity Functions} \label{subsec:obs_LF}
In Figure~\ref{fig:rate_vs_M}, we compare our LF to the $r$- or $R$-band LFs from \citet{li11a}, \citet{perley20}, and \citet{sharon-kushnir22}. We convert these LFs to the $V$ band using
\begin{equation} \label{eq:r_to_v}
    M_V = M_{0,V} + \frac{b_V}{b_r} \left( M_r - M_{0,r} \right)
\end{equation}
where the values of $M_{0,V} = -19.12 \pm 0.01\, \mathrm{mag}$, $M_{0,r} = -19.03 \pm 0.01\, \mathrm{mag}$, $b_V = 0.95 \pm 0.11$, and $b_r = 1.02 \pm 0.11$ are from Table~9 of \citet[][]{folatelli10}. \citet{sharon-kushnir22} compute the observed LF for the LOSS survey shown in Figure~\ref{fig:rate_vs_M} by scaling the rates to match the total BTS rate at the peak of the LF. The two most luminous bins in our LF only include SNe~Ia-CSM, which are not present in the LF from \citet{perley20}, causing the steep decline in rate at the luminous end.
The low-luminosity end of our LF is consistent with the LF from \citet{perley20}, however, we see a significant difference for the higher luminosities. In their rate calculations, \citet{perley20} used an average completeness correction factor of $f_{\mathrm{rec}} = 0.6$ for the recovery efficiency of transients. As seen in Figure~\ref{fig:comp_vs_mag}, the completeness is a function of absolute magnitude because more luminous SNe are brighter for longer. The use of a single correction factor would result in an underestimate of the completeness for high luminosity SNe leading to an overestimate of the associated rates. 
The rates from \citet{perley20} also do not seem to account for the time dilation of the rates, but this effect would be less important since the redshifts are modest.
The magnitude shifts predicted by Eq.~\ref{eq:r_to_v} are uncertain by roughly $0.1\,\mathrm{mag}$, which is not large enough to explain the factor of $\sim 5$ difference in the LFs near $M_V = -19.5\,\mathrm{mag}$.

A larger effect that can explain the difference is host-galaxy extinction. We can make the luminous end of the LFs agree if we have an average extra contribution to the right side of Eq.~\ref{eq:r_to_v} of $\left(R_V - \frac{b_V}{b_r}R_r\right) E(V-r)$. If we use a typical Galactic extinction law with $R_V=3.1$ and $R_r=2.3$ \citep{Cardelli89}, then a mean extinction of $E(V-r) \approx 0.21\,\mathrm{mag}$ provides the necessary shift. If we use an empirical extinction law derived from observations of SNe~Ia but of uncertain interpretation, such as $R_V=1.74$ and $R_r=0.89$ from \citet{folatelli10}, then a mean extinction of $E(V-r) \approx 0.22\,\mathrm{mag}$ again provides the necessary shift. These are roughly consistent with the estimates of $E(B-V) \approx 0.1$ and $E(B-V) \approx 0.17$ found by \citet{sharon-kushnir22} and \citet{Burns14}, respectively.

\citet{sharon-kushnir22} attempted to correct for the host-galaxy extinction and obtain an intrinsic LF by using the colour stretch $s_{gr}$ from the $g$- and $r$-band light curves with the luminosities calibrated using the CSP SNe Ia sample \citep{Contreras10,Stritzinger11,Krisciunas17,Burns18}. The resulting LF is more sharply peaked probably due to ignoring the scatter in the peak magnitude -- $s_{gr}$ relation, which leads to a narrower distribution in magnitudes. The \citet{perley20} and \citet{sharon-kushnir22} rates are based on the same ZTF BTS sample but \citet{sharon-kushnir22} find higher rates. An additional contribution to the rate difference may be that \citet{sharon-kushnir22} use a shorter survey duration compared to the BTS analysis. Using the longer duration of \citet{perley20} lowers the \citet{sharon-kushnir22} rate to $2.56^{+0.58}_{-0.46}\,\rateunits$, which is within $10\%$ of the rate from \citet{perley20}.

\begin{figure}                          
	\includegraphics[width=\columnwidth]{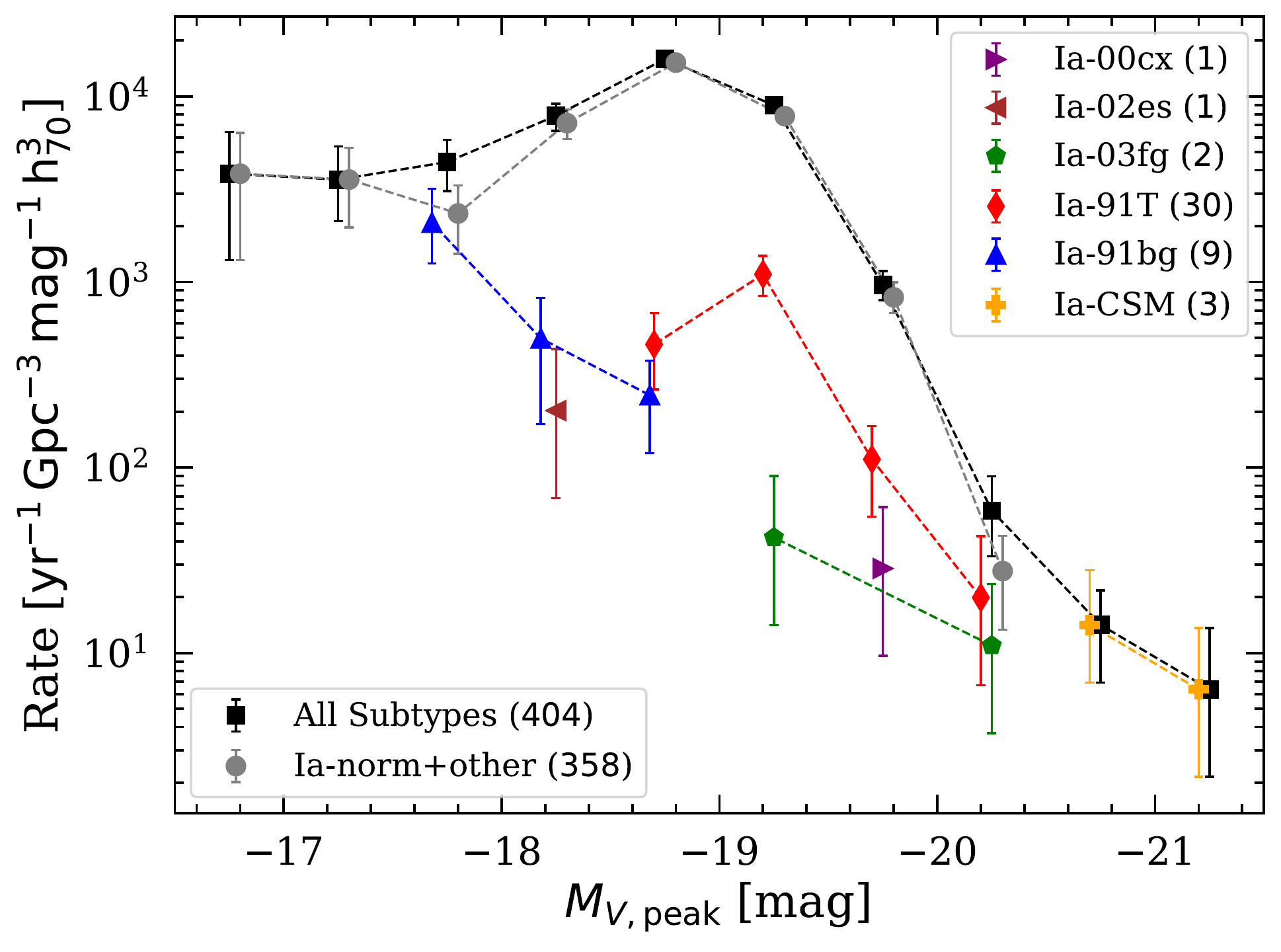}
    \caption{Luminosity functions for various subtypes of SNe Ia. The total LF for all subtypes is shown in black. Ia-norm+other includes normal SNe Ia as well as all SNe not classified into a specific subtype, only having the classification of `Ia.' For clarity, LFs for Ia-91T and Ia-CSM are shifted to the left by $0.05\,\mathrm{mag}$, Ia-91bg by $0.07\,\mathrm{mag}$, and Ia-norm+other to the right by $0.05\,\mathrm{mag}$. The sample size for each subtype is given in parentheses.}
    \label{fig:rate_vs_M_per_type}
\end{figure}

\begin{table*}
\centering
\caption{Luminosity Functions and Total Rates for Each Major subtype of SNe Ia.}
\label{tab:rates_subtype}
% \resizebox{\columnwidth}{!}
{
\renewcommand{\arraystretch}{1.5}
\begin{tabular}{ccccccccc}
\hline
\hline
    $M_{\mathrm{\textit{V},peak}}$ & $R_{\mathrm{all}}$ & $R_{\mathrm{Ia-norm}}$ & $R_{\mathrm{Ia-91T}}$ & $R_{\mathrm{Ia-91bg}}$ & $R_{\mathrm{Ia-CSM}}$ & $R_{\mathrm{Ia-03fg}}$ & $R_{\mathrm{Ia-02es}}$ & $R_{\mathrm{Ia-00cx}}$\\

    [mag] & \multicolumn{8}{c}{$[\mathrm{yr}^{-1}\, \mathrm{Gpc}^{-3}\, \mathrm{mag}^{-1}\, h^{3}_{70}]$} \\
\hline
    $[-16.5, -17.0]$ & $4^{+3}_{-3} \times 10^{3}$          & $4^{+3}_{-3} \times 10^{3}$          & ---                               & --- & --- & --- & --- & --- \\
    $[-17.0, -17.5]$ & $3.6^{+1.8}_{-1.4} \times 10^{3}$    & $3.6^{+1.8}_{-1.4} \times 10^{3}$    & ---                               & --- & --- & --- & --- & --- \\
    $[-17.5, -18.0]$ & $4.4^{+1.3}_{-1.3} \times 10^{3}$    & $2.3^{+1.0}_{-1.0} \times 10^{3}$    & ---                               & $2.1^{+1.0}_{-0.9} \times 10^{3}$ & --- & --- & --- & --- \\
    $[-18.0, -18.5]$ & $8.0^{+1.4}_{-1.3} \times 10^{3}$    & $7.2^{+1.3}_{-1.4} \times 10^{3}$    & ---                               & $5.0^{+0.3}_{-0.3} \times 10^{2}$ & --- & --- & $2.0^{+2.0}_{-1.3} \times 10^{2}$ & --- \\
    $[-18.5, -19.0]$ & $1.59^{+0.14}_{-0.13} \times 10^{4}$ & $1.52^{+0.14}_{-0.13} \times 10^{4}$ & $5^{+2}_{-2} \times 10^{2}$       & $2.5^{+1.5}_{-1.3} \times 10^{2}$ & --- & --- & --- & --- \\
    $[-19.0, -19.5]$ & $9.0^{+0.8}_{-0.8} \times 10^{3}$    & $7.8^{+0.7}_{-0.7} \times 10^{3}$    & $1.1^{+0.3}_{-0.2} \times 10^{3}$ & --- & --- & $40^{+50}_{-30}$ & --- & --- \\
    $[-19.5, -20.0]$ & $9.7^{+1.7}_{-1.6} \times 10^{2}$    & $8.3^{+1.7}_{-1.5} \times 10^{2}$    & $1.1^{+0.6}_{-0.5} \times 10^{2}$ & --- & --- & --- & --- & $29^{+30}_{-19}$ \\
    $[-20.0, -20.5]$ & $60^{+30}_{-30}$                     & $28^{+15}_{-15}$                     & $20^{+20}_{-13}$                  & --- & --- & $11^{+12}_{-7}$ & --- & --- \\
    $[-20.5, -21.0]$ & $14^{+8}_{-7}$                       & ---                                  & ---                               & --- & $14^{+8}_{-7}$ & --- & --- & --- \\
    $[-21.0, -21.5]$ & $6^{+7}_{-4}$                        & ---                                  & ---                               & --- & $6^{+7}_{-4}$ & --- & --- & --- \\
\hline
     & \multicolumn{8}{c}{Total Rates $[\mathrm{yr}^{-1}\, \mathrm{Gpc}^{-3}\, h^{3}_{70}]$} \\
\hline
    $[-16.5, -21.5]$ & $2.28^{+0.20}_{-0.20} \times 10^{4}$  & $2.04^{+0.20}_{-0.19} \times 10^{4}$  & $8.5^{+1.6}_{-1.7} \times 10^{2}$  & $1.4^{+0.5}_{-0.5} \times 10^{3}$ & $10^{+7}_{-7}$ & $30^{+20}_{-20}$ & $1.0^{+1.0}_{-0.8} \times 10^{2}$ & $14^{+16}_{-9}$ \\
    $[-17.0, -21.5]$ & $2.09^{+0.15}_{-0.15} \times 10^{4}$  & $1.85^{+0.14}_{-0.14} \times 10^{4}$  & $8.5^{+1.6}_{-1.7} \times 10^{2}$  & $1.4^{+0.5}_{-0.5} \times 10^{3}$ & $10^{+7}_{-7}$ & $30^{+20}_{-20}$ & $1.0^{+1.0}_{-0.8} \times 10^{2}$ & $14^{+16}_{-9}$ \\
    $[-17.5, -21.5]$ & $1.91^{+0.12}_{-0.12} \times 10^{4}$  & $1.67^{+0.11}_{-0.11} \times 10^{4}$  & $8.5^{+1.6}_{-1.7} \times 10^{2}$  & $1.4^{+0.5}_{-0.5} \times 10^{3}$ & $10^{+7}_{-7}$ & $30^{+20}_{-20}$ & $1.0^{+1.0}_{-0.8} \times 10^{2}$ & $14^{+16}_{-9}$ \\
\hline
\end{tabular}} \\
\begin{flushleft}
\noindent \textbf{Note:} Results shown in this table are for a host-galaxy extinction of $0\,\mathrm{mag}$. The column of $R_{\mathrm{Ia-norm}}$ includes normal SNe Ia as well as SNe not classified into a specific subtype.
\end{flushleft}
\end{table*}

Because of our high spectroscopic completeness, we are also able to compute the LF of SNe Ia for different subtypes for the first time. Specifically, the subtypes include Ia-norm (\& other), Ia-91bg, Ia-91T, Ia-CSM, Ia-03fg, Ia-02es, and Ia-00cx. Ia-norm (\& other) includes the normal SNe Ia as well as the SNe Ia that were not classified into a subtype. In total there are 358 Ia-norm (\& other), 9 Ia-91bg, 30 Ia-91T, 3 Ia-CSM, 2 Ia-03fg, 1 Ia-02es, and 1 Ia-00cx.
Figure \ref{fig:rate_vs_M_per_type} shows the LF for each subtype of SNe Ia and Table~\ref{tab:rates_subtype} gives the rates. The LF of SNe Ia-91bg peaks at a fainter magnitude ($M_{\mathrm{\textit{V},peak}} \sim -17.75\,\mathrm{mag}$) than that of SNe Ia-91T ($M_{\mathrm{\textit{V},peak}} \sim -19.25\,\mathrm{mag}$), illustrating that SNe Ia-91bg are intrinsically fainter than SNe Ia-91T. From the total rates $R_{\mathrm{Ia-91bg}}$ and $R_{\mathrm{Ia-91T}}$, we also note that SNe Ia-91bg are intrinsically more common than SNe Ia-91T, but since SNe Ia-91bg are faint and more difficult to classify, we do not observe as many of them. Finally, as expected, we see that only SNe Ia-CSM contribute to the two most luminous bins and are the only SNe~Ia more luminous than $M_{\mathrm{\textit{V},peak}} = -21\, \mathrm{mag}$ due to their CSM interaction.

The rates and LFs for each subtype shown here serve as a lower limit, since it is likely that some SNe~Ia classified without a subtype fall in this category. \citet{li11a} found the observed fractions of subtypes of SNe~Ia in a volume-limited sample; SNe~Ia-norm are $\sim70\%$ of the total, SNe~Ia-91bg are $\sim15\%$, and SNe~Ia-91T are $\sim9\%$. In comparison, we find a larger fraction of 89\% for SNe~Ia-norm (\& other), and lower fractions of 6\% for SNe~Ia-91bg, and 4\% for SNe~Ia-91T. The fractions for the rarer subtypes are $0.4\%$ for SNe~Ia-02es, $0.1\%$ for SNe~Ia-03fg, $0.06\%$ for SNe~Ia-00cx, and $0.04\%$ for SNe~Ia-CSM.  We note that our SNe~Ia-CSM rate is consistent with the recent ZTF-BTS rate determination of $\sim0.02\% - 0.2\%$ \citep{Sharma23}. For our sample, the classification of SNe~Ia-norm (\& other) includes SNe~Ia that are not classified into any specific subtype in addition to the `normal' SNe~Ia, thus artificially increasing the fraction of `normal' SNe~Ia and reducing the fractions of subtypes. A more careful, uniform spectroscopic classification is needed in the future to classify SNe~Ia into appropriate subtypes.

\subsection{Correcting for Host-Galaxy Extinction} \label{subsec:host_ext_LF}
Correcting for the host-galaxy extinction in each SN requires multi-band light curves, which we do not have for this SN sample. \citet{sharon-kushnir22} used the colour stretch to correct for host-galaxy extinction and found a mean selective extinction of $E(B-V) \approx 0.1$. From their reanalysis of the LOSS data, they found a mean extinction of $A_V \sim 0.5\,\mathrm{mag}$ with a tail out to $A_V \sim 2 - 3\,\mathrm{mag}$. Similarly, the ZTF-BTS sample of host galaxies shows a mean extinction of $A_r \sim 0.25\,\mathrm{mag}$ and $A_g \sim 0.25\,\mathrm{mag}$, both tailing out to values of $\sim 2\,\mathrm{mag}$.

\begin{figure}
	\includegraphics[width=\columnwidth]{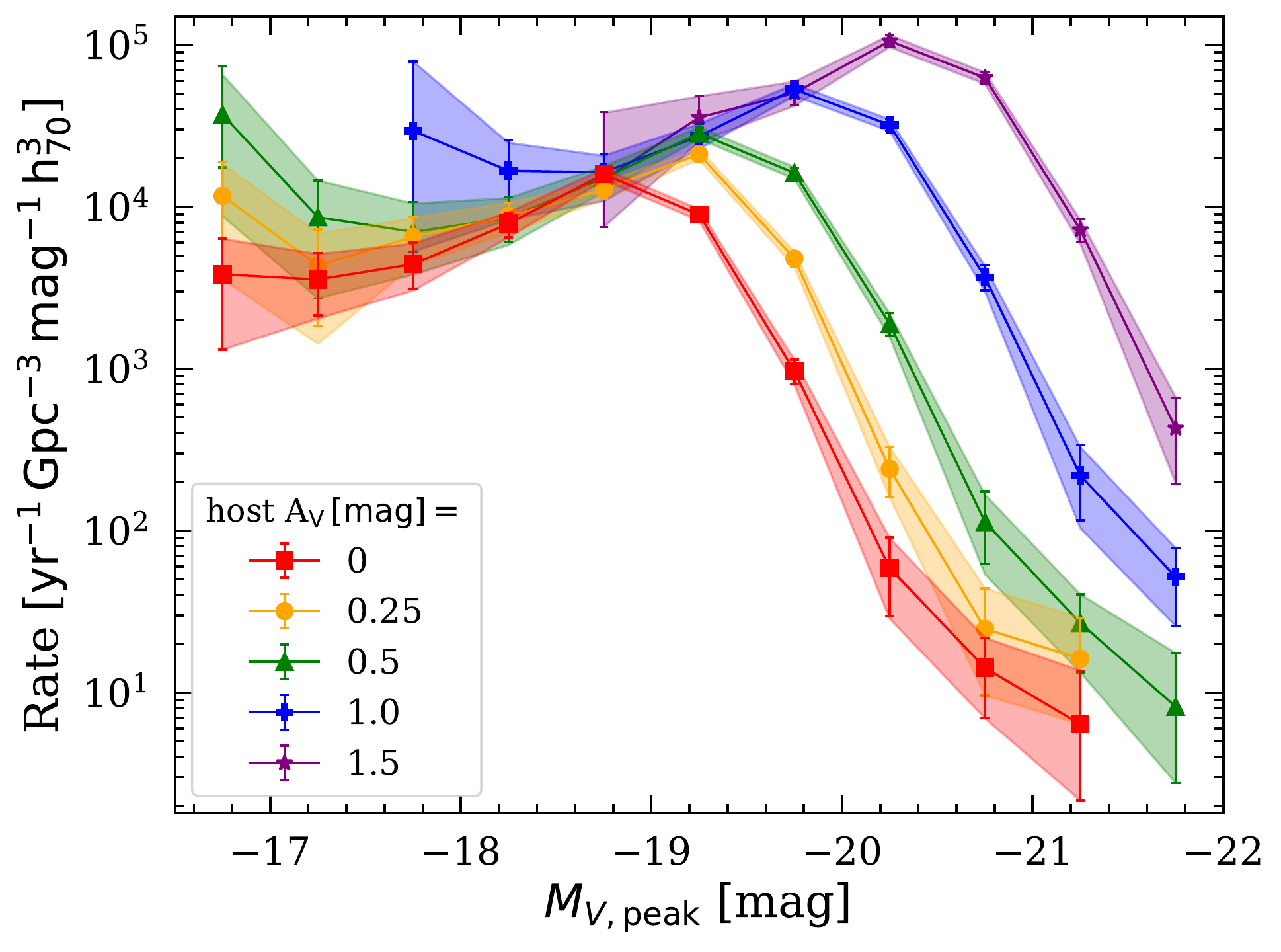}
    \caption{Luminosity functions for various assumed values of average host-galaxy extinction $A_{V}$.}
    \label{fig:rate_vs_M_host_av}
\end{figure}

Here, we provide LFs as a function of host-galaxy extinction values which can be weighted to provide any `desired' correction. We show the LFs for different values of host-galaxy extinction $A_{V} = $ ($0.0,\,0.25,\,0.5,\,1.0$, and $1.5\,\mathrm{mag}$) in Figure~\ref{fig:rate_vs_M_host_av}. The overall effect is nearly identical to simply adding a constant extinction to all SNe leading to a shift in the rates of $10^{0.6 A_V}$ modulo small second-order effects. As outlined in Section~\ref{subsec:obs_LF}, the LFs in two different filters can also be used to estimate the host-extinction. Assuming that our LF and the LF from \citet{perley20} are both correct, they must agree with each other after accounting for the host-galaxy extinction term when converting from one filter to another. Using this method, we estimate a mean host extinction of $E(V-r) \approx 0.2\,\mathrm{mag}$.

\section{Summary}
\label{sec:summary}
We use a nearly spectroscopically complete catalogue of SNe Ia from ASAS-SN \citep{holoien17a, holoien17b, holoien17c, holoien19} to estimate the volumetric SN Ia rate in the local universe ($z<0.08$). We start by refitting all ASAS-SN light curves using the $V$-band SN Ia templates from \citet{nugent02} to improve the estimates of the peak magnitudes (see Section~\ref{sec:data}). We use randomly drawn light curves and injected SNe to estimate completeness corrections as a function of the peak apparent and absolute magnitudes of SNe. After weighing each observed SN according to its peak apparent and absolute magnitudes, we compute the volumetric rate (see Eq.~\ref{eq:rate}).

After considering the effect of various limiting cuts (see Section~\ref{subsec:vol_rate}), we use a sample of 404 SNe Ia at a median redshift of $z_{\mathrm{med}}=0.024$ for our standard results. The choice of $|b_\mathrm{lim}| = 15\degr$ and $m_{\mathrm{\textit{V},lim}} = 17\,\mathrm{mag}$ provides a balance between statistical and systematic uncertainties for our sample. The standard sample yields a total volumetric rate of $R_{\mathrm{tot}} = 2.28^{+0.20}_{-0.20} \rateunits$. This rate is in agreement with rates from other studies at low redshifts \citep{cappellaro99,dilday10,li11b,frohmaier19,perley20,Srivastav22}, but has smaller uncertainties. 

We construct the observed LF and compare it with LFs from \citet{li11a}, \citet{perley20}, and \citet{sharon-kushnir22}. We use a completeness correction as a function of both apparent and absolute magnitudes, rather than an average completeness. Assuming that our LF and that from \citep{perley20} are both correct, we estimate a mean host extinction of $E(V-r) \approx 0.2\,\mathrm{mag}$ based on the magnitude shift between the LFs in the different filters ($V$ and $r$). We also compute, for the first time, LFs for the major subtypes of SNe Ia, finding that the less luminous SNe Ia-91bg are more numerous than the more luminous SNe Ia-91T (see Section~\ref{subsec:obs_LF}). Finally, we provide the LFs corrected for several values of average host-galaxy extinction (see Section~\ref{subsec:host_ext_LF}).

In the upcoming papers, we plan on updating these SNe Ia rates using the newer $g$-band ASAS-SN data from \citet{Neumann23} spanning SNe discovered from 2018 to 2020. The ASAS-SN $g$-band observations are sensitive to objects $\sim$1\,mag fainter than the $V$-band data and will roughly double the sample size. This will allow improving the estimates for the rates and LFs. Such a large sample will also allow us to obtain the rates as a function of other parameters such as galaxy type, local SN environments, or separation of an SN from its host. We will employ a more careful uniform spectroscopic classification to classify SNe~Ia into appropriate subtypes and improve the subtype rates. We will also be able to calculate rates for other populations, in particular, the core-collapse (CC) SNe and the tidal disruption events (TDEs). This paper is the first in a series that will measure rates and LFs for SNe Ia, CC SNe, TDEs, and other transients from ASAS-SN.

\section*{Acknowledgements}
We thank Federica Chiti, Jason Hinkle, Willem Hoogendam, and Michael Tucker for helpful discussions.

We thank Las Cumbres Observatory and its staff for their continued support of ASAS-SN. ASAS-SN is funded in part by the Gordon and Betty Moore Foundation through grants GBMF5490 and GBMF10501 to the Ohio State University, and also funded in part by the Alfred P. Sloan Foundation grant G-2021-14192. Development of ASAS-SN has been supported by NSF grant AST-0908816, the Mt. Cuba Astronomical Foundation, the Center for Cosmology and AstroParticle Physics at the Ohio State University, the Chinese Academy of Sciences South America Center for Astronomy (CAS-SACA), and the Villum Foundation. 

We wish to extend our special thanks to those of Hawaiian ancestry
on whose sacred mountains of Maunakea and Haleakal\=a, we
are privileged to be guests. Without their generous hospitality, the
observations presented herein would not have been possible.

Support for TJ was provided by NASA through the NASA Hubble Fellowship grant HF2-51509 awarded by the Space Telescope Science Institute, which is operated by the Association of Universities for Research in Astronomy, Inc., for NASA, under contract NAS5-26555.
KZS is supported by the 2022 Guggenheim Fellowship, and his stay at the UCSB KITP was supported in part by the National Science Foundation under NSF Grant No.\ PHY-1748958.
JFB was supported by NSF Grant No.\ PHY-2012955.
SD acknowledges support by the National Natural Science Foundation of China (Grant No. 12133005).

%%%%%%%%%%%%%%%%%%%%%%%%%%%%%%%%%%%%%%%%%%%%%%%%%%
\section*{Data Availability}
The data used in this paper are available in a machine-readable format from the online journal as supplementary material. A portion is shown in Table~\ref{tab:SNeIa} for guidance regarding its form and content.



\bibliographystyle{mnras}
\bibliography{IaRates} % if your bibtex file is called example.bib

%\begin{thebibliography}{99}
%\end{thebibliography}

%%%%%%%%%%%%%%%%%%%%%%%%%%%%%%%%%%%%%%%%%%%%%%%%%%

%%%%%%%%%%%%%%%%% APPENDICES %%%%%%%%%%%%%%%%%%%%%
\appendix
\section{Rates and Luminosity Functions for Non-Zero Host-Galaxy Extinction}
\begin{table*}
\centering
\caption{Luminosity Functions and Total Rates for a Mean Host-Galaxy Extinction of $A_V = 0.25\,\mathrm{mag}$.}
\label{tab:rates_AV0.25}
% \resizebox{\columnwidth}{!}
{
\renewcommand{\arraystretch}{1.5}
\begin{tabular}{ccccccccc}
\hline
\hline
    $M_{\mathrm{\textit{V},peak}}$ & $R_{\mathrm{all}}$ & $R_{\mathrm{Ia-norm}}$ & $R_{\mathrm{Ia-91T}}$ & $R_{\mathrm{Ia-91bg}}$ & $R_{\mathrm{Ia-CSM}}$ & $R_{\mathrm{Ia-03fg}}$ & $R_{\mathrm{Ia-02es}}$ & $R_{\mathrm{Ia-00cx}}$\\

    [mag] & \multicolumn{8}{c}{$[\mathrm{yr}^{-1}\, \mathrm{Gpc}^{-3}\, \mathrm{mag}^{-1}\, h^{3}_{70}]$} \\
\hline
    $[-16.5, -17.0]$ & $1.2^{+0.7}_{-0.8} \times 10^{4}$    & $1.2^{+0.7}_{-0.8} \times 10^{4}$    & ---                               & --- & --- & --- & --- & --- \\
    $[-17.0, -17.5]$ & $4^{+3}_{-2} \times 10^{3}$          & $4^{+3}_{-2} \times 10^{3}$          & ---                               & --- & --- & --- & --- & --- \\
    $[-17.5, -18.0]$ & $7^{+2}_{-2} \times 10^{3}$          & $5.1^{+1.9}_{-1.9} \times 10^{3}$    & ---                               & $1.5^{+1.2}_{-0.8} \times 10^{3}$ & --- & --- & --- & --- \\
    $[-18.0, -18.5]$ & $8.8^{+2.1}_{-1.8} \times 10^{3}$    & $6.6^{+1.6}_{-1.5} \times 10^{3}$    & ---                               & $1.9^{+1.0}_{-0.9} \times 10^{3}$ & --- & --- & $2.8^{+3.2}_{-1.9} \times 10^{2}$ & --- \\
    $[-18.5, -19.0]$ & $1.28^{+0.16}_{-0.16} \times 10^{4}$ & $1.22^{+0.16}_{-0.16} \times 10^{4}$ & ---                               & $6^{+3}_{-3} \times 10^{2}$       & --- & --- & --- & --- \\
    $[-19.0, -19.5]$ & $2.12^{+0.15}_{-0.15} \times 10^{4}$ & $1.95^{+0.15}_{-0.14} \times 10^{4}$ & $1.7^{+0.4}_{-0.4} \times 10^{3}$ & --- & --- & --- & --- & --- \\
    $[-19.5, -20.0]$ & $4.8^{+0.5}_{-0.5} \times 10^{3}$    & $4.2^{+0.5}_{-0.5} \times 10^{3}$    & $5.2^{+1.7}_{-1.5} \times 10^{2}$ & --- & --- & $60^{+70}_{-40}$ & --- & $40^{+40}_{-30}$ \\
    $[-20.0, -20.5]$ & $2.4^{+0.8}_{-0.8} \times 10^{2}$    & $1.8^{+0.7}_{-0.7} \times 10^{3}$    & $60^{+30}_{-30}$                  & --- & --- & --- & --- & --- \\
    $[-20.5, -21.0]$ & $25^{+19}_{-15}$                     & ---                                  & ---                               & --- & $10^{+11}_{-6}$ & $15^{+17}_{-10}$ & --- & --- \\
    $[-21.0, -21.5]$ & $16^{+10}_{-10}$                     & ---                                  & ---                               & --- & $16^{+10}_{-10}$ & --- & --- & --- \\
\hline
     & \multicolumn{8}{c}{Total Rates $[\mathrm{yr}^{-1}\, \mathrm{Gpc}^{-3}\, h^{3}_{70}]$} \\
\hline
    $[-16.5, -21.5]$ & $3.5^{+0.4}_{-0.4} \times 10^{4}$    & $3.0^{+0.4}_{-0.4} \times 10^{4}$    & $1.1^{+0.2}_{-0.2} \times 10^{3}$ & $2.0^{+0.7}_{-0.7} \times 10^{3}$ & $13^{+8}_{-8}$ & $40^{+30}_{-30}$ & $1.4^{+1.6}_{-1.0} \times 10^{2}$ & $20^{+20}_{-15}$ \\
    $[-17.0, -21.5]$ & $2.9^{+0.2}_{-0.2} \times 10^{4}$    & $2.6^{+0.2}_{-0.2} \times 10^{4}$    & $1.1^{+0.2}_{-0.2} \times 10^{3}$ & $2.0^{+0.7}_{-0.7} \times 10^{3}$ & $13^{+8}_{-8}$ & $40^{+30}_{-30}$ & $1.4^{+1.6}_{-1.0} \times 10^{2}$ & $20^{+20}_{-15}$ \\
    $[-17.5, -21.5]$ & $2.72^{+0.19}_{-0.18} \times 10^{4}$ & $2.39^{+0.17}_{-0.17} \times 10^{4}$ & $1.1^{+0.2}_{-0.2} \times 10^{3}$ & $2.0^{+0.7}_{-0.7} \times 10^{3}$ & $13^{+8}_{-8}$ & $40^{+30}_{-30}$ & $1.4^{+1.6}_{-1.0} \times 10^{2}$ & $20^{+20}_{-15}$ \\
\hline
\end{tabular}} \\
\begin{flushleft}
\noindent \textbf{Note:} The column of $R_{\mathrm{Ia-norm}}$ includes normal SNe Ia as well as SNe not classified into a specific subtype.
\end{flushleft}
\end{table*}

\begin{table*}
\centering
\caption{Luminosity Functions and Total Rates for a Mean Host-Galaxy Extinction of $A_V = 0.5\,\mathrm{mag}$.}
\label{tab:rates_AV0.5}
% \resizebox{\columnwidth}{!}
{
\renewcommand{\arraystretch}{1.5}
\begin{tabular}{ccccccccc}
\hline
\hline
    $M_{\mathrm{\textit{V},peak}}$ & $R_{\mathrm{all}}$ & $R_{\mathrm{Ia-norm}}$ & $R_{\mathrm{Ia-91T}}$ & $R_{\mathrm{Ia-91bg}}$ & $R_{\mathrm{Ia-CSM}}$ & $R_{\mathrm{Ia-03fg}}$ & $R_{\mathrm{Ia-02es}}$ & $R_{\mathrm{Ia-00cx}}$\\

    [mag] & \multicolumn{8}{c}{$[\mathrm{yr}^{-1}\, \mathrm{Gpc}^{-3}\, \mathrm{mag}^{-1}\, h^{3}_{70}]$} \\
\hline
    $[-16.5, -17.0]$ & $3.7^{+0.3}_{-0.3} \times 10^{4}$    & $3.7^{+0.3}_{-0.3} \times 10^{4}$    & ---                               & --- & --- & --- & --- & --- \\
    $[-17.0, -17.5]$ & $9^{+6}_{-5} \times 10^{3}$          & $9^{+6}_{-5} \times 10^{3}$          & ---                               & --- & --- & --- & --- & --- \\
    $[-17.5, -18.0]$ & $7^{+3}_{-3} \times 10^{3}$          & $7^{+3}_{-3} \times 10^{3}$          & ---                               & --- & --- & --- & --- & --- \\
    $[-18.0, -18.5]$ & $9^{+3}_{-3} \times 10^{3}$          & $4.5^{+2.1}_{-1.7} \times 10^{3}$    & ---                               & $4.0^{+2.0}_{-1.9} \times 10^{3}$ & --- & --- & --- & --- \\
    $[-18.5, -19.0]$ & $1.5^{+0.3}_{-0.2} \times 10^{4}$    & $1.4^{+0.3}_{-0.3} \times 10^{4}$    & ---                               & $9^{+6}_{-6} \times 10^{2}$       & --- & --- & $4^{+4}_{-3} \times 10^{2}$ & --- \\
    $[-19.0, -19.5]$ & $2.9^{+0.3}_{-0.2} \times 10^{4}$    & $2.7^{+0.2}_{-0.2} \times 10^{4}$    & $8.2^{+0.4}_{-0.3} \times 10^{2}$ & $5^{+2}_{-2} \times 10^{2}$       & --- & --- & --- & --- \\
    $[-19.5, -20.0]$ & $1.62^{+0.13}_{-0.13} \times 10^{4}$ & $1.41^{+0.12}_{-0.11} \times 10^{4}$ & $2.0^{+0.5}_{-0.4} \times 10^{3}$ & --- & --- & $80^{+90}_{-60}$ & --- & --- \\
    $[-20.0, -20.5]$ & $1.9^{+0.4}_{-0.3} \times 10^{3}$    & $1.6^{+0.3}_{-0.3} \times 10^{3}$    & $2.2^{+1.1}_{-1.0} \times 10^{2}$ & --- & --- & --- & --- & $60^{+60}_{-40}$ \\
    $[-20.5, -21.0]$ & $1.1^{+0.6}_{-0.5} \times 10^{2}$    & $50^{+30}_{-30}$                     & $40^{+40}_{-30}$                  & --- & --- & $21^{+24}_{-14}$ & --- & --- \\
    $[-21.0, -21.5]$ & $27^{+14}_{-13}$                     & ---                                  & ---                               & --- & $27^{+14}_{-13}$ & --- & --- & --- \\
    $[-21.5, -22.0]$ & $8^{+9}_{-5}$                        & ---                                  & ---                               & --- & $8^{+9}_{-5}$ & --- & --- & --- \\
\hline
     & \multicolumn{8}{c}{Total Rates $[\mathrm{yr}^{-1}\, \mathrm{Gpc}^{-3}\, h^{3}_{70}]$} \\
\hline
    $[-16.5, -22.0]$ & $6.2^{+1.6}_{-1.6} \times 10^{4}$    & $5.3^{+1.5}_{-1.5} \times 10^{4}$    & $1.5^{+0.3}_{-0.3} \times 10^{3}$ & $2.7^{+1.0}_{-1.0} \times 10^{3}$ & $18^{+11}_{-11}$ & $50^{+40}_{-40}$ & $2.0^{+2.0}_{-1.5} \times 10^{2}$ & $30^{+30}_{-20}$\\
    $[-17.0, -22.0]$ & $4.3^{+0.4}_{-0.4} \times 10^{4}$    & $3.8^{+0.4}_{-0.4} \times 10^{4}$    & $1.5^{+0.3}_{-0.3} \times 10^{3}$ & $2.7^{+1.0}_{-1.0} \times 10^{3}$ & $18^{+11}_{-11}$ & $50^{+40}_{-40}$ & $2.0^{+2.0}_{-1.5} \times 10^{2}$ & $30^{+30}_{-20}$\\
    $[-17.5, -22.0]$ & $3.9^{+0.3}_{-0.3} \times 10^{4}$    & $3.4^{+0.3}_{-0.3} \times 10^{4}$    & $1.5^{+0.3}_{-0.3} \times 10^{3}$ & $2.7^{+1.0}_{-1.0} \times 10^{3}$ & $18^{+11}_{-11}$ & $50^{+40}_{-40}$ & $2.0^{+2.0}_{-1.5} \times 10^{2}$ & $30^{+30}_{-20}$\\
\hline
\end{tabular}} \\
\begin{flushleft}
\noindent \textbf{Note:} The column of $R_{\mathrm{Ia-norm}}$ includes normal SNe Ia as well as SNe not classified into a specific subtype.
\end{flushleft}
\end{table*}

\begin{table*}
\centering
\caption{Luminosity Functions and Total Rates for a Mean Host-Galaxy Extinction of $A_V = 1.0\,\mathrm{mag}$.}
\label{tab:rates_AV1.0}
% \resizebox{\columnwidth}{!}
{
\renewcommand{\arraystretch}{1.5}
\begin{tabular}{ccccccccc}
\hline
\hline
    $M_{\mathrm{\textit{V},peak}}$ & $R_{\mathrm{all}}$ & $R_{\mathrm{Ia-norm}}$ & $R_{\mathrm{Ia-91T}}$ & $R_{\mathrm{Ia-91bg}}$ & $R_{\mathrm{Ia-CSM}}$ & $R_{\mathrm{Ia-03fg}}$ & $R_{\mathrm{Ia-02es}}$ & $R_{\mathrm{Ia-00cx}}$\\

    [mag] & \multicolumn{8}{c}{$[\mathrm{yr}^{-1}\, \mathrm{Gpc}^{-3}\, \mathrm{mag}^{-1}\, h^{3}_{70}]$} \\
\hline
    $[-17.5, -18.0]$ & $3^{+5}_{-3} \times 10^{4}$          & $3^{+5}_{-3} \times 10^{4}$          & ---                               & --- & --- & --- & --- & --- \\
    $[-18.0, -18.5]$ & $1.7^{+0.9}_{-0.7} \times 10^{4}$    & $1.7^{+0.9}_{-0.7} \times 10^{4}$    & ---                               & --- & --- & --- & --- & --- \\
    $[-18.5, -19.0]$ & $1.6^{+0.5}_{-0.5} \times 10^{4}$    & $8.7^{+0.4}_{-0.4} \times 10^{3}$    & ---                               & $8^{+4}_{-3} \times 10^{3}$       & --- & --- & --- & --- \\
    $[-19.0, -19.5]$ & $2.7^{+0.5}_{-0.5} \times 10^{4}$    & $2.5^{+0.5}_{-0.4} \times 10^{4}$    & ---                               & $1.7^{+1.2}_{-1.1} \times 10^{3}$ & --- & --- & $7^{+8}_{-5} \times 10^{2}$ & --- \\
    $[-19.5, -20.0]$ & $5.3^{+0.5}_{-0.4} \times 10^{4}$    & $5.1^{+0.4}_{-0.4} \times 10^{4}$    & $1.6^{+0.7}_{-0.6} \times 10^{3}$ & $8^{+4}_{-4} \times 10^{2}$       & --- & --- & --- & --- \\
    $[-20.0, -20.5]$ & $3.2^{+0.3}_{-0.3} \times 10^{4}$    & $2.8^{+0.3}_{-0.2} \times 10^{4}$    & $4.0^{+0.9}_{-0.9} \times 10^{3}$ & --- & --- & $1.6^{+1.8}_{-1.1} \times 10^{2}$ & --- & --- \\
    $[-20.5, -21.0]$ & $3.7^{+0.6}_{-0.6} \times 10^{3}$    & $3.1^{+0.6}_{-0.6} \times 10^{3}$    & $4.2^{+0.2}_{-0.2} \times 10^{2}$ & --- & --- & --- & --- & $1.1^{+1.2}_{-0.7} \times 10^{2}$ \\
    $[-21.0, -21.5]$ & $2.2^{+1.2}_{-1.0} \times 10^{2}$    & $1.0^{+0.6}_{-0.5} \times 10^{2}$    & $70^{+80}_{-50}$                  & --- & --- & $40^{+50}_{-30}$ & --- & --- \\
    $[-21.5, -22.0]$ & $50^{+30}_{-30}$                     & ---                                  & ---                               & --- & $50^{+30}_{-30}$ & --- & --- & --- \\
\hline
     & \multicolumn{8}{c}{Total Rates $[\mathrm{yr}^{-1}\, \mathrm{Gpc}^{-3}\, h^{3}_{70}]$} \\
\hline
    $[-17.5, -22.0]$ & $8.6^{+2.8}_{-2.7} \times 10^{4}$ & $7.9^{+2.8}_{-2.6} \times 10^{4}$ & $3.0^{+0.6}_{-0.6} \times 10^{3}$ & $5.1^{+1.9}_{-1.9} \times 10^{3}$ & $26^{+13}_{-13}$ & $1.0^{+0.8}_{-0.8} \times 10^{2}$ & $4^{+4}_{-3} \times 10^{2}$ & $50^{+60}_{-30}$ \\
\hline
\end{tabular}} \\
\begin{flushleft}
\noindent \textbf{Note:} The column of $R_{\mathrm{Ia-norm}}$ includes normal SNe Ia as well as SNe not classified into a specific subtype.
\end{flushleft}
\end{table*}

\begin{table*}
\centering
\caption{Luminosity Functions and Total Rates for a Mean Host-Galaxy Extinction of $A_V = 1.5\,\mathrm{mag}$.}
\label{tab:rates_AV1.5}
% \resizebox{\columnwidth}{!}
{
\renewcommand{\arraystretch}{1.5}
\begin{tabular}{ccccccccc}
\hline
\hline
    $M_{\mathrm{\textit{V},peak}}$ & $R_{\mathrm{all}}$ & $R_{\mathrm{Ia-norm}}$ & $R_{\mathrm{Ia-91T}}$ & $R_{\mathrm{Ia-91bg}}$ & $R_{\mathrm{Ia-CSM}}$ & $R_{\mathrm{Ia-03fg}}$ & $R_{\mathrm{Ia-02es}}$ & $R_{\mathrm{Ia-00cx}}$\\

    [mag] & \multicolumn{8}{c}{$[\mathrm{yr}^{-1}\, \mathrm{Gpc}^{-3}\, \mathrm{mag}^{-1}\, h^{3}_{70}]$} \\
\hline
    $[-18.5, -19.0]$ & $1.5^{+2.4}_{-1.0} \times 10^{4}$    & $1.5^{+2.4}_{-1.0} \times 10^{4}$    & ---                               & --- & --- & --- & --- & --- \\
    $[-19.0, -19.5]$ & $3.6^{+1.2}_{-1.1} \times 10^{4}$    & $1.9^{+0.9}_{-0.7} \times 10^{4}$    & ---                               & $1.7^{+0.8}_{-0.8} \times 10^{4}$ & --- & --- & --- & --- \\
    $[-19.5, -20.0]$ & $5.1^{+0.9}_{-0.9} \times 10^{4}$    & $4.6^{+0.9}_{-0.7} \times 10^{4}$    & ---                               & $3^{+2}_{-2} \times 10^{3}$       & --- & --- & $1.3^{+1.5}_{-0.9} \times 10^{3}$ & --- \\
    $[-20.0, -20.5]$ & $1.06^{+0.09}_{-0.08} \times 10^{5}$ & $1.01^{+0.09}_{-0.08} \times 10^{5}$ & $3.1^{+1.4}_{-1.2} \times 10^{3}$ & $1.7^{+0.9}_{-0.8} \times 10^{3}$ & --- & --- & --- & --- \\
    $[-20.5, -21.0]$ & $6.3^{+0.5}_{-0.5} \times 10^{4}$    & $5.5^{+0.5}_{-0.4} \times 10^{4}$    & $7.7^{+1.7}_{-1.7} \times 10^{3}$ & --- & --- & $3^{+4}_{-2} \times 10^{2}$ & --- & --- \\
    $[-21.0, -21.5]$ & $7.2^{+1.3}_{-1.3} \times 10^{3}$    & $6.2^{+1.3}_{-1.1} \times 10^{3}$    & $8^{+4}_{-4} \times 10^{2}$       & --- & --- & --- & --- & $2.1^{+2.4}_{-1.4} \times 10^{2}$ \\
    $[-21.5, -22.0]$ & $4.3^{+2.1}_{-1.9} \times 10^{2}$    & $2.0^{+1.0}_{-1.0} \times 10^{2}$    & $1.5^{+1.7}_{-1.0} \times 10^{2}$ & --- & --- & $80^{+90}_{-50}$ & --- & --- \\
\hline
     & \multicolumn{8}{c}{Total Rates $[\mathrm{yr}^{-1}\, \mathrm{Gpc}^{-3}\, h^{3}_{70}]$} \\
\hline
    $[-17.5, -22.0]$ & $1.33^{+0.15}_{-0.15} \times 10^{5}$ & $1.16^{+0.15}_{-0.14} \times 10^{5}$ & $5.9^{+1.1}_{-1.2} \times 10^{3}$  & $1.1^{+0.4}_{-0.4} \times 10^{4}$ & --- & $2.0^{+1.6}_{-1.6} \times 10^{2}$ & $7^{+7}_{-5} \times 10^{2}$ & $1.1^{+1.1}_{-0.8} \times 10^{2}$ \\
\hline
\end{tabular}} \\
\begin{flushleft}
\noindent \textbf{Note:} The column of $R_{\mathrm{Ia-norm}}$ includes normal SNe Ia as well as SNe not classified into a specific subtype.
\end{flushleft}
\end{table*}

% If you want to present additional material which would interrupt the flow of the main paper,
% it can be placed in an Appendix which appears after the list of references.

%%%%%%%%%%%%%%%%%%%%%%%%%%%%%%%%%%%%%%%%%%%%%%%%%%

% Don't change these lines
\bsp	% typesetting comment
\label{lastpage}
\end{document}